Article type: Review

# Theory, preparation, properties and catalysis application in 2D Graphynes-Based Materials

*Ning Zhang[1]\*, Jiayu Wu[1], Taoyuan Yu[1], Jiaqi Lv[1], He Liu[1] and Xiping Xu[1]*

\*Corresponding Author Email: zhangning@cust.edu.cn

[1]College of Photoelectrical Engineering, Changchun University of Science and Technology, Changchun 130022, China

## Abstract

Carbon has three hybridization forms of sp-, $sp^2$- and $sp^3$-, and the combination of different forms can obtain different kinds of carbon allotropes, such as diamond, carbon nanotubes, fullerene, graphynes (GYs) and graphdiyne (GDY). Among them, the GDY molecule is a single-layer two-dimensional (2D) planar structure material with highly π-conjugation formed by sp- and $sp^2$-hybridization. GDY has a carbon atom ring composed of benzene ring and acetylene, which makes GDY have a uniformly distributed pore structure. In addition, GDY planar material have some slight wrinkles, which makes GDY have better self-stability than other 2D planar materials. The excellent properties of GDY make it attract the attention of researcher. Therefore, GDY is widely used in chemical catalysis, electronics, communications, clean energy and composite materials. This paper summarizes the recent progress of GDY research, including structure, preparation, properties and application of GDY in the field of catalysts.

Key words: 2D materials, catalysis application, graphyne, property

## 1. Introduction

With the development of some emerging photonic applications such as flexible sensing and biomonitoring [1-3], two dimensional materials may offer potential to put up a better integrated photonic system, because two dimensional materials possess several unique characteristics [4-15]. Except single two dimensional materials, van der Waals heterostructure with an intriguing phenomenon of interlayer excitons has drawn much attention recent years, which offers a potential to achieve NIR lasers [16-23]. Synthesis, separation of new and different dimension of carbon allotropes is the focus of research over the past two hundred and thirty years. Scientists have discovered the three-dimensional fullerenes, one dimensional carbon nanotubes and two-dimensional graphene new carbon allotropes. These materials have become the frontier and hot issue of international academic research. Additionally, graphene, demonstrating high saturation current density, good stability at high temperature, ultrafast heating and cooling speed, mature technology in fiber optics [24-33], is an ideal material for thermal emitters. Graphdiyne (GDY) is a new carbon allotrope, which is composed of sp and $sp^2$ hybrid carbon network and has atomic porosity [34,35]. Due to its special structure, GDY has completely different properties and applications compared with other carbon allotropes [36,37] such as carbon nanotubes [38-41] and graphene [31,42-44]. Graphdiyne has rich carbon bonds, high π-conjugation, wide plane spacing, uniformly dispersed pore configurations and controllable electronic structures. This feature gives it great potential in gas separation [45,46], catalysis [47-49] and energy-related fields [50-56]. In 1987, Baughman calculated theoretically that the carbyne material formed by sp and $sp^2$ hybrid carbon, namely graphyne, could exist stably [34]. Since graphdiyne was proposed, many famous research groups in the world have carried out exploration and research [57-59]. However, due to the high energy and flexibility of the sp bond, graphdiyne has not been successfully prepared. Until 2010, Li Yuliang's research group made an important breakthrough in the preparation of graphdiyne, and successfully synthesized a large-area (3.61 $cm^2$) graphdiyne thin film with two-dimensional structure on the surface of copper sheet by chemical method [35,60]. The research progress in recent decades is shown in Figure 1, among which more than half of the articles related to graphyne have discussed GDY. Therefore, this article will also focus on GDY.

Graphdiyne is an all-carbon molecule with 2D planar network structure formed by the conjugated connection of six benzene rings by diacetylene bond, and its unique molecular configuration is determined by the bonding mode of the hybrid state of sp and $sp^2$. As the most stable diacetylene allotrope [57], the direct band gap of GDY is 0.46 eV. Tthe intrinsic electron mobility is predicted to be as high as $10^5$ $cm^2$ $V^{-1}$ $s^{-1}$ at room temperature, which is about an order of magnitude higher than the hole mobility [61]. Until 2010, Li et al. successfully synthesized large area graphdiyne thin film by chemical method for the first time in the world [35,60], laying a foundation for the transition from theoretical research to experimental test of graphdiyne. GDY materials come in a variety of forms, such as thin films, nanospheres, nanotube arrays and nanowires, which can be achieved by special synthetic methods [62,63]. With excellent electrical, optical and photoelectrical properties, GDY has potential important applications in electronics, energy, catalysis, photovoltaics, water repair and other fields [64-73]. The research on the manufacture of high quality GDY has developed rapidly and gradually formed a new research hotspot and field. The research on the manufacture of high quality GDY has developed rapidly and gradually formed a new research hotspot and field. As we discussed before, two dimensional materials including graphene, TMDCs, BP have exhibited various advantages for photonic and opto-electronic applications than their bulk counterparts [74-85]. However, the polycrystalline or non-morphological GDY obtained at present is different from the ideal GDY, which results in the performance of GDY is worse than that of theoretical prediction [86]. In addition, there is still no breakthrough in the preparation of controllable macroquantity, structure characterization and property regulation of GDY based on its important basis.

In this review, the theory, preparation, properties and catalytic applications of GDY are reviewed and summarized. The second part mainly introduces the structure and bond length of GDY, as well as the theoretical methods used to calculate the prediction model. The third part describes the mechanical, electronic and optical properties of GDY. Finally, in the fourth part, the latest advances in the use of GDY for catalysis are reviewed. In the future work, we will focus on the preparation of a large number of high-quality GDY slices and methods to control the aggregation and assembly of GDY and GY to form a hierarchical structure.

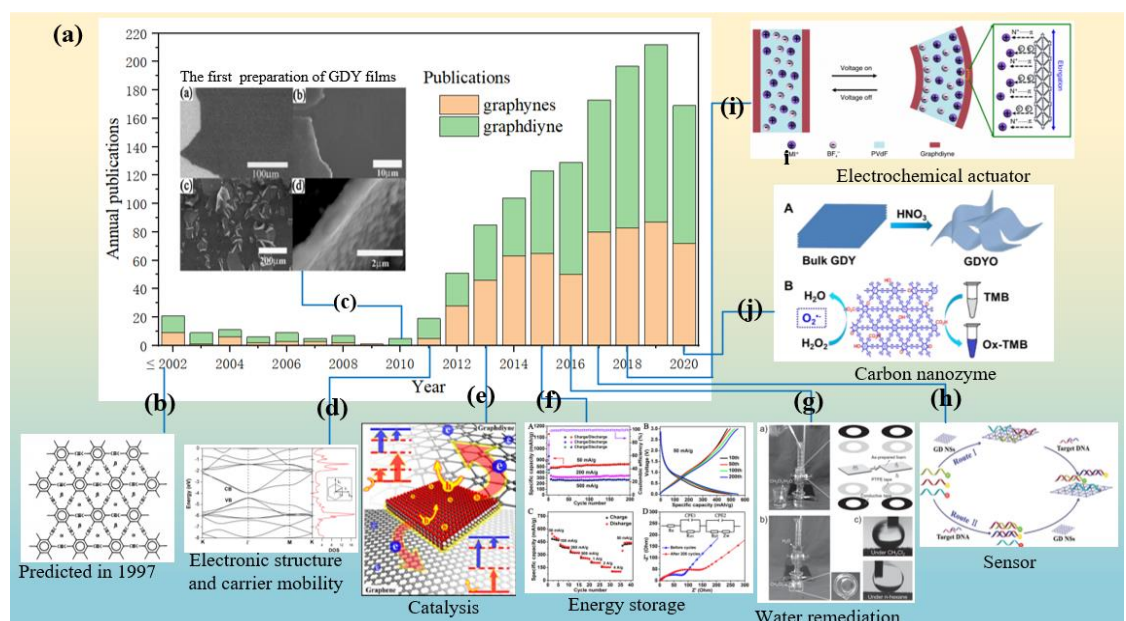

Figure 1. Recent advances in graphynes and graphdiyne research (1987–2020 July), data queried from Web of Science[TM]. (a) Journal publications and worldwide patents issued per year. (b) The structure of graphdiyne was calculated by Baughman. (c) The first successful preparation of GDY films. Reprinted with permission from ref. [60]. Copyright 2010 The Royal Society of Chemistry. (d) Calculated electronic structure of GDY. Reprinted with permission from ref. [61]. Copyright 2011 American Chemical Society. (e–j) Selected GDY-related applications, including the following reports: Wang et al., ACS nano, 2013 (reprinted with permission from ref. [64]. Copyright 2013 American Chemical Society), Zhang et al., Chem. Commun., 2015 (reprinted with permission from ref. [66]. Copyright 2015 The Royal Society of Chemistry), Zhang et al., Adv. Mater., 2016 (reprinted with permission from ref. [67]. Copyright 2016 Wiley-VCH), Wang et al., Adv. Mater., 2017 (reprinted with permission from ref. [68]. Copyright 2017 Wiley-VCH), Chen et al., Nat. Commun., 2018 (reprinted with permission from ref. [71]. Copyright 2018 Nature Publishing Group) and Mao et al., Chem. Commun., 2020 (reprinted with permission from ref. [73]. Copyright 2020 The Royal Society of Chemistry).

## 2. Theory and calculation of graphyne

### 2.1 Structure and bond length

In 1968, Baughman, a theoretical chemist, calculated theoretically that the carbyne material formed by $sp^2$ and sp hybrid carbon could be stable [87]. In 1997, Haley proposed that graphdiyne (GDY) was the most stable carbon allotrope containing diacetylene linkers [57]. The structure of graphyne can be seen as a two-dimensional hexagonal carbon ring network connected by acetylene bonds (sp hybridization), as shown in Figure 2(a). This graphyne structure is referred to in many literature as $\gamma$-graphyne to separate it from other members of the graphyne family [88,89]. However, the original notation in Bowman's article is still followed here and is called graphyne [34]. Graphyne has a hexagonal symmetry similar to graphene, but the length of the acetylene bond can vary. This results in a graphyne-$n$ structure, where $n$ represents the number of −C≡C− bonds in the bond (Figure 2(a)). The first experimentally successful member of the graphyne family is graphdiyne (graphyne-2) [60]. The reason why the structural stability of graphyne-$n$ is much weaker than that of graphyne is that the cohesion energy of the acetylene bond is decreased when it is inserted into the carbon network [34]. In graphyne-$n$ structure, graphyne is considered to be the most stable. However, the binding energy decreases as the length of the acetylene bond increases [34].

The lattice constant $a_n$ of graphyne-$n$ can be obtained:

$$a_n = a_{\text{graphyne}} + (n-1)\Delta a_{\text{acetylene}} \tag{1}$$

where $a_{\text{graphyne}}$ is the lattice constant of graphyne, and $\Delta a_{\text{acetylene}}$ is the length of the acetylene bond in the structure [90]. In the formula, the optimal value of $a_{\text{graphyne}}$ is between 6.86Å and 6.90 Å [34,58,91,92], and the value of $\Delta a_{\text{acetylene}}$ is approximately 2.58Å [58,93]. The different C-C bond types of graphyne-$n$ are due to the co-existence of sp and sp$^2$ hybridization. There are three types of carbon-carbon bonds in graphenine molecules, namely benzene ring aromatic bonds (the theoretical length of the bond is 0.142 nm), the single bond connecting C=C and C≡C (the theoretical length is 0.134nm) and C≡C bond (the theoretical length is 0.120 nm) [58,90,94,95]. The research results of carbon-carbon bond length [90] are shown in **Table 1**. Due to the weak coupling between the acetylene unit and the benzene ring, these single bonds are shortened and aromatic bonds are extended compared with typical single bonds and aromatic bonds [96], which reflects the hybridization effect of sp and sp$^2$ carbon atoms. Average bond length is usually used to quantitatively determine lattice spacing. First principles and molecular dynamics (MD) [97] calculations show that the lattice spacing increases uniformly as the size of graphyne increases [58]. For example, for each additional acetylene connection unit, the lattice spacing increased at a regular rate of about 0.266 nm, while the quantum level analysis showed an increase of about 0.258 nm (with slight variations due to differences in bond lengths obtained by using different atomic orbital methods) [58]. These results show that prolonging the acetylene connection unit does not result in large structural changes.

**Table 1** Compare calculated equilibrium bond lengths (Å) [90]

| Work | Aromatic | Single | Triple | Note(s) |
| --- | --- | --- | --- | --- |
| Cranford and Buchler [98] | 1.48-1.50 | 1.46-1.48 | 1.18-1.19 | MD, ReaxFF potential; extended graphynes |
| Baughman et al.[34] | 1.428 | 1.421 | 1.202 | MNDO; canonical(1987); graphynes |
| Yang and Xu [97] | 1.405-1.406 | 1.341-1.396 | 1.239-1.240 | MD, AIREBO potential; extended graphynes$^a$ |
| Narita et al. [58] | 1.419 | 1.401 | 1.221 | DFT, LSDA; extended graphynes |
| Bai et al. [99] | 1.440 | 1.341-1.400 | 1.239 | DFT, GGA-PBE; graphdiyne only$^a$ |
| Mirnezhad et al. [100] | 1.423 | 1.404 | 1.219 | DFT, GGA-PBE; graphyne only |
| Peng et al. [101] | 1.426 | 1.407 | 1.223 | KS-DFT; graphyne only |
| Pei [102] | 1.431 | 1.337-1.395 | 1.231 | VASP, GGA-PBE; graphdiyne only$^a$ |

$^a$ Range in single bond length due to differentiation between interior single bonds (connecting two sp$^1$ carbons and exterior single bonds connecting sp$^1$ with and sp$^2$ carbon).

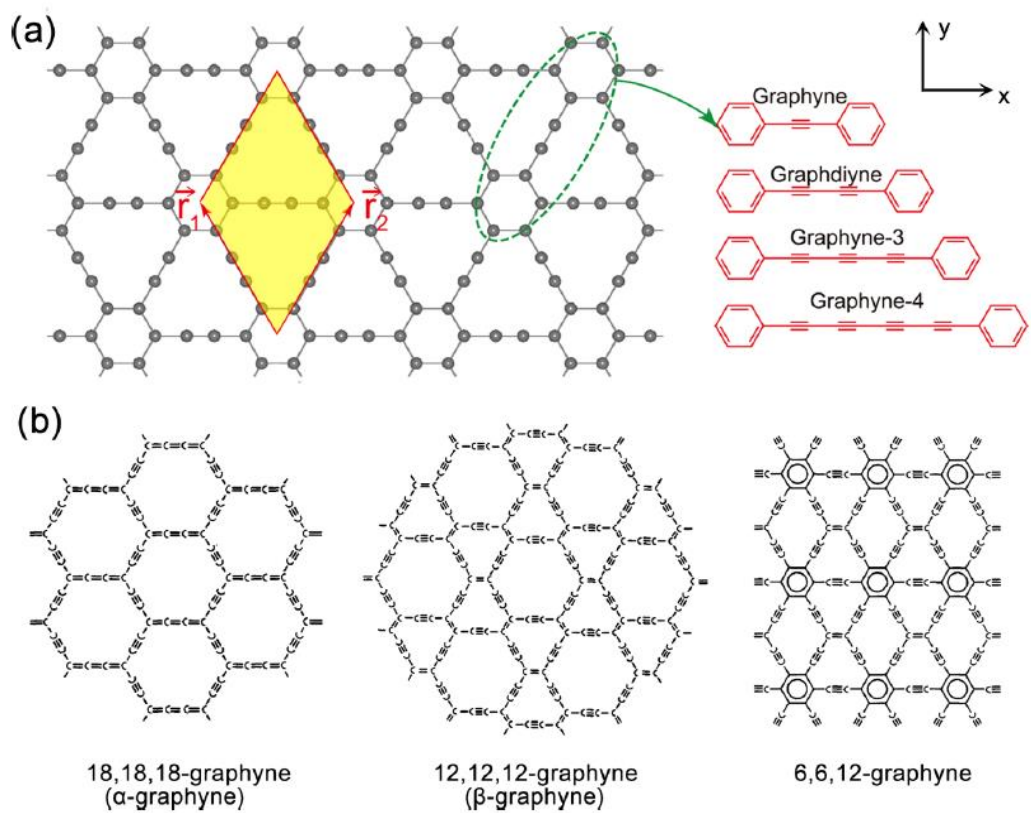

Figure 2. (a) Illustration of the graphyne-n structures. Reprinted with permission from ref. [40] Copyright 2013 American Chemical Society.(b) Structures of $\alpha$-graphyne, $\beta$-graphyne, and 6,6,12-graphyne. Reprinted from ref. [103] with the permission of AIP Publishing.

Graphyne family members can have many other structures besides graphyne-$n$ structure. This is because $sp^2$ and sp carbon atoms can form two-dimensional networks with different arrays. These structures can be expressed as: $\alpha$, $\beta$, $\gamma$-graphyne [34]. Here, $\alpha$ and $\beta$ represent the number of C atoms in the minimum and subminimum rings connected by the acetylene bond, and $\gamma$ is the number of C atoms in the third ring connected to $\beta$ by the acetylene bond. The proportions of sp and $sp^2$ carbon atoms in $\alpha-\text{graphyne}$, $\beta-\text{graphyne}$ and $\gamma$-graphyne structures are different and their symmetry are different, as shown in Figure 2(b). The structures of 18,18,18-graphyne, 12,12,12-graphyne, and 6,6,12-grapyne are given respectively. As can be seen from the Figure 2(b), the symmetry of the first two structures is hexagonal, while 6,6,12 graphyne has rectangular symmetry. 18,18,18-graphyne and 12,12,12-graphyne are commonly referred to as $\alpha$-graphyne and $\beta$-graphyne [104].

It is found that graphenine C-C bond can lead to greater structural variability of graphenine than graphyne, which is conducive to the formation of curved nanowires and nanotubes. Moreover, it is predicted that graphenine has low generation energy and high thermal stability. Compared with graphyne and some other $sp^2$ graphyne isoallotropes, acetylene (diacetylene) acts as a connecting unit in these two-dimensional carbon networks, which relatively reduces its stability. Baughman et al. [59] used the energy of each atom to evaluate the relative stability of various graphyne, and predicted the high-temperature stability of graphyne with 12.4 kJ/mol per carbon atom. Qiao et al. [99] calculated the energy of graphyne diacetylene as 0.803eV/ atom (relative to graphyne). The corresponding values for diamond, graphite, (6,6)-carbon nanotubes, C60 and carbin are approximately 0.002, 0.00, 8,0.114, 0.364 and 1.037eV/ atom, respectively. The benzene ring and C≡C bond is a large triangular ring with 18 carbon atoms, with a pore diameter of about 0.25 nm. The hybridized acetylene bond and benzene ring of sp and $sp^2$ give graphyne a two-dimensional single-atom plane configuration. In order to maintain the stability of the structure, certain folds will be formed in the process of infinite plane expansion and extension. At the same time, the layers of two-dimensional graphyne are piled up by van der Waals force and π-π interaction to form three-dimensional layered structure. The triangular holes of 18 carbon atoms form uniformly dispersed three-dimensional channels in space. Therefore, graphyne has the characteristics of both two-dimensional planar materials and three-dimensional porous materials.

Carbon nanotubes, graphene and GDY are three representative carbon element materials. Graphene is very thin, but its strength is really high which is 200 times higher than steel. Graphene is considered a potential substitute for silicon material, which is due to graphene can greatly increase the computing speed of traditional processors. Graphene can be used as a raw material for transparent electronic products due to its unique light absorption rate. Carbon nanotubes were discovered earlier than graphene, but carbon nanotubes can be regarded as a curled structure of graphene. Analogously, carbon nanotubes also have high strength, light weight and excellent conductivity. GDY was discovered later than the above two materials and it shows great potential for applications. There are more carbon chemical bonds, larger conjugated system and excellent chemical stability in GDY. When GDY is used as a base material for electrocatalysts, it can improve catalytic activity and stability effectively. GDY can be used as a storage material in lithium batteries due to its excellent conductivity and capacity. The discovery of GDY has a great impact on carbon chemistry, and it can be applied in many fields.

**2.2 Calculation method**

In order to simulate the properties of graphyne and its family members (GFMs), several theoretical methods for calculation are described. Density functional theory (DFT) is a quantum mechanical method to study the electronic structure of multi-electron systems. The DFT proved that the properties of the system depend on the electron density of the multi-electron system [105]. The objective of density functional theory is to replace the complex multi-electron wave function with the electron density as the basic quantity of research. Because the multi-electron wave function has 3N variables (N is the number of electrons, each electron contains three spatial variables), and the electron density is only a function of three variables, thus greatly reducing the processing difficulty. The most common application of DFT is achieved by the Kohn-Sham method. In the framework of Kohn-sham DFT, the most difficult multibody problem (resulting from electron interactions in an external electrostatic potential) is reduced to the problem of electrons moving in an effective potential field without interacting. The effective potential field includes the external potential field and the effects of coulomb interactions between electrons, such as exchange and correlation [106]. It is difficult to deal with exchange correlation in KS DFT. There is no exact method to solve the exchange correlation energy $E_{xc}$. The simplest approximation is the local density approximation (LDA approximation) [106,107]. LDA approximately uses uniform electron gas to calculate the exchange energy of the system (the exchange energy of uniform electron gas can be precisely solved), while the correlation energy is treated by fitting the free electron gas. Snice LDA is based on the ideal uniform electron gas model, the electron density of actual atomic and molecular systems is far from uniform. To further improve the calculation accuracy, the inhomogeneity of electron density should be considered.

This is generally accomplished by introducing the gradient of electron density into the exchange correlation function, that is, the generalized gradient approximates (GGA) [108]. However, both the LDA and GGA significantly underestimated the band gap in their calculations [109].

GW approximation is used to calculate the self energy in multi-body systems. The multi-body perturbation theory based on Green's function provides a strict theoretical framework for describing the properties of ground and excited states of materials. Green's function relies on the exchange of associated energy, which satisfies a complex set of integral-differential equations called Hedin equations [110]. GW method is based on the multi-body perturbation theory of the self-energy operator based on the shielded coulomb effect and is the most accurate first-principle method to describe the quasiparticle electron excitation properties of the extended system [111]. GW approximately uses the interaction between Green function G and W with shield to expand the system energy and calculate $\sum_{xc}$ [111,112]. The GW calculation is very large, and the GW approximation for even simple binary compounds is considerable. Therefore, it is hoped to reduce the calculation amount without affecting the calculation accuracy. So far, there is no simplified GW theory applicable to the system used. Most simplified theories apply only to semiconductors and insulators. At present, GW approximation has achieved great success in material property calculation, and some models have been established. But these models are not very reliable. Because the band gap values in these models are ideal, but the band structure is not so satisfactory. The main problem faced by GW computing is to construct a simple self-energy model and apply it to complex systems without sacrificing accuracy, and at the same time, the model satisfies energy correlation and non-local properties.

Semi-empirical tight-binding method is used to simulate the performance of GFMS. The tight-binding approximation is an empirical method for energy band structure calculation. In 1928, Bloch proposed the inbound approximation, which expands the electronic states in a crystal by linear combinations of atomic orbitals [113]. It extends the single electron wave function of the system based on atomic orbital. The Hamiltonian matrix elements between these atomic orbitals are treated as tunable parameters and fit the results of experiments or first-principle calculations, and then the eigenvalues and eigenstates are calculated by diagonalizing the Hamiltonian matrix [114]. The tight-binding model gives a reasonable description of the electron occupied state of any type of crystal (metals, semiconductors, and insulators), which is much cheaper than the DFT calculation. However, the fitting parameters of the tight-binding method have strong system dependence, so its transferability is limited.

The Non-equilibrium Green's function (NEGF) method is suitable for the energy transmission problem caused by atomic vibration (phonons) [115-118]. It is mainly used in small junction systems. In this method, the simulation system consists of three parts. Two semi-infinite leads act as electrons or hot baths, and they are connected by a central conductor region. Electron or phonon transmission is calculated based on green's function of the central region and the self-energy of the lead wires describing the interaction between the lead centers. NEGF preserves quantum mechanical effects such as tunneling and diffraction, and can well describe the quantum heat transport and spin thermoelectric properties of low-Dimensional nano materials. But NEGF is not suitable for large equipment due to high computing costs.

## 3. Preparation and properties

## 3.1 Preparation

3.1.1 Films

Before preparing GDY, the first step was to understand the GDY structure by theoretical prediction, as shown in Figure 3. At first, the structure of GYs was proposed by Baughman et al.[34]. Some other groups were also discussed to create GYs network. Next, Haley et al. tried to get GDY, but failed [57]. Finally, the substructure of GDY was obtained through a series of studies. In recent years, Li et al. proposed a new scheme to successfully prepare GDY through the original cross-coupling reaction on the basis of conversion method, metal-catalyzed cross-coupling method and template-assisted synthesis method [60].

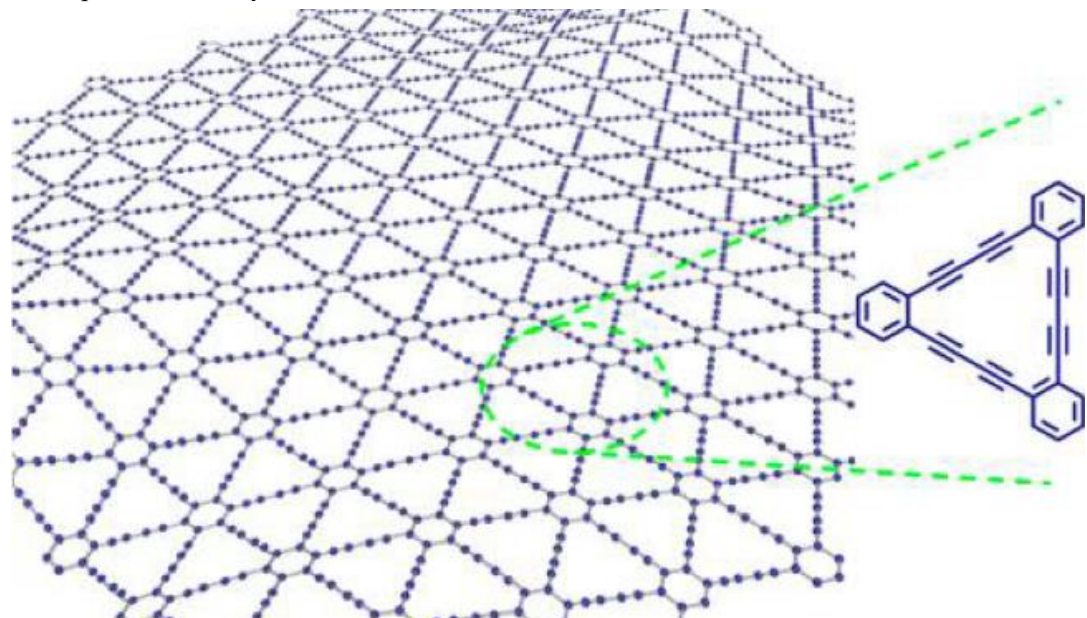

Figure 3 Schematic representation of the GDY structure. Reproduced with permission from ref.34 Copyright 2018 Chemical Reviews.

Hexaethynylbenzene (HEB) was selected as precursor to prepare GDY film. Copper foil was reaction substrate, it was also the polymerization catalyst, as shown in Figure 4 [60]. The copper foil was submerged in the reaction solution, through the provision of copper ion catalyzed coupling reaction. The intermediate process produced Cu acetylated material $C-Cu^{2+}$ bond. Finally, uniform multilayer GDY membrane can be obtained through a series of reactions.

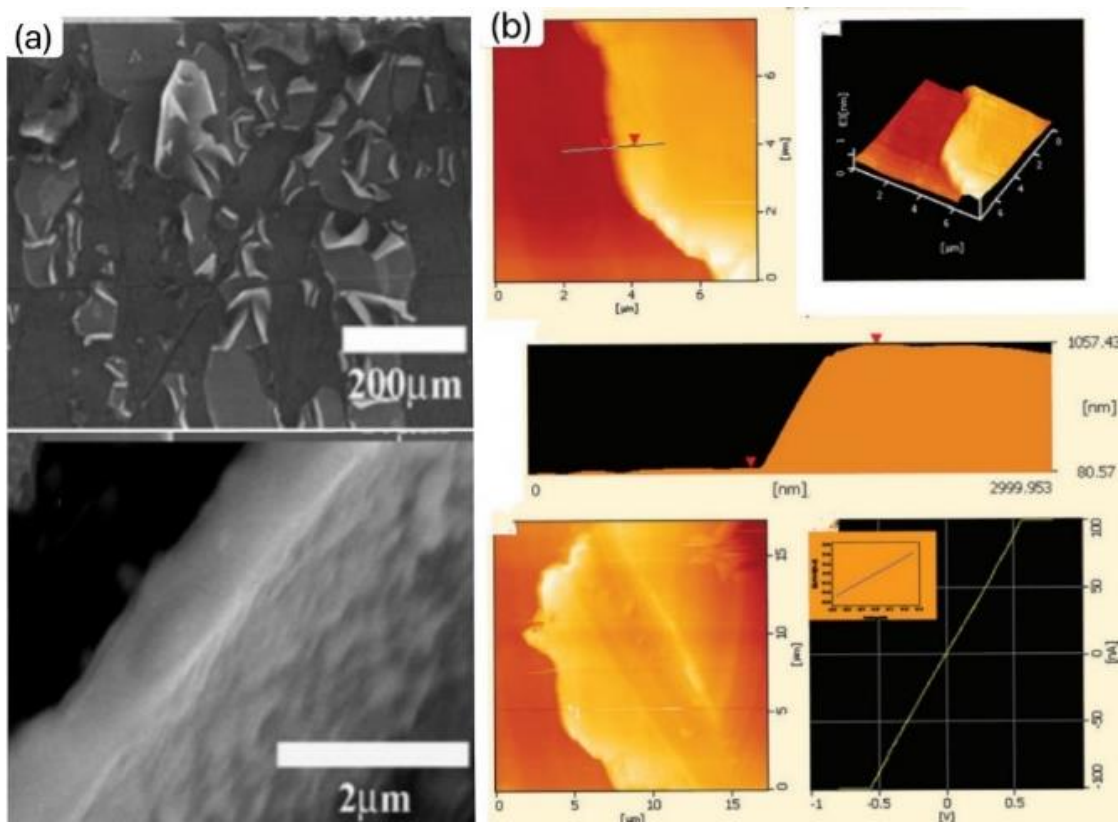

Figure 4 (a) Structures of precursor HEB and GDY, scanning electron microscopy (SEM) and photographic images, and (b) atomic force microscopy (AFM) images of GDY films with its I−V curve. Reproduced with permission from ref. 60. Copyright 2010 Royal Society of Chemistry

By controlling the weight of the GDY powder and the moving position of the heating, different layers of GDY could be prepared with the gas-liquid-solid process [119]. In the heating engineering, part of the zinc oxide reduced to zinc was used as GDY membrane catalysts. The distance of different GDY was about 0.36 mm when observed through a high-resolution electron microscope. Gas liquid solid method for preparation of high conductivity GDY film provided a new train of thought.

However, almost all methods require copper as a substrate, and the high cost of copper substrate severely limits its large-scale production. Through the other process, single layer and multilayer GDY could be prepared on Ag substrate. As shown in Figure 5, this method could prepare a small amount of GDY. The prepared thin film had semiconductor properties [120].

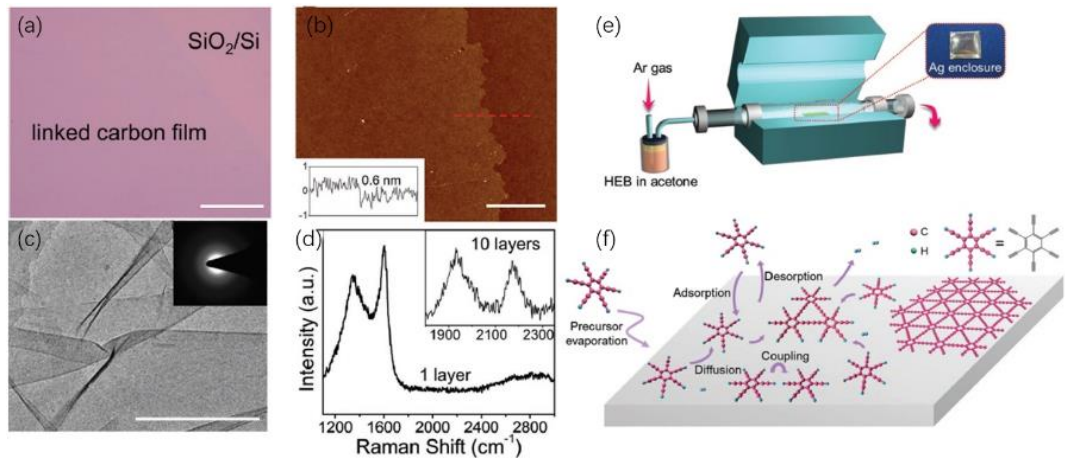

Figure 5 (a) Illustration of the CVD system for the GDY growth on silver surface. (b) Illustration of the growth process. (c) Optical microscope image of the as-prepared GDY film on the surface of SiO2/Si substrate. (d) Selected-area electron diffraction (SAED) pattern and TEM image of the film. (e) Height profile and AFM image of the film on SiO2/Si substrate. (f) Raman spectrum of the as-grown single-layer and 10-layer films. Reproduced with permission from ref. 120. Copyright 2017 John Wiley and Sons.

Subsequently, Zhang et al. successfully synthesized GDY film [121], as shown in Figure 6. Atomic repeated structures of graphene and GDY interacted when they were put together. This was a key step to prepare ultra-thin GDY. In addition, the above methods could also prepare similar thin films. Recently, the thickness and conductivity of the GDY structure were adjusted and optimized by controlling the release rate of copper ions [122].

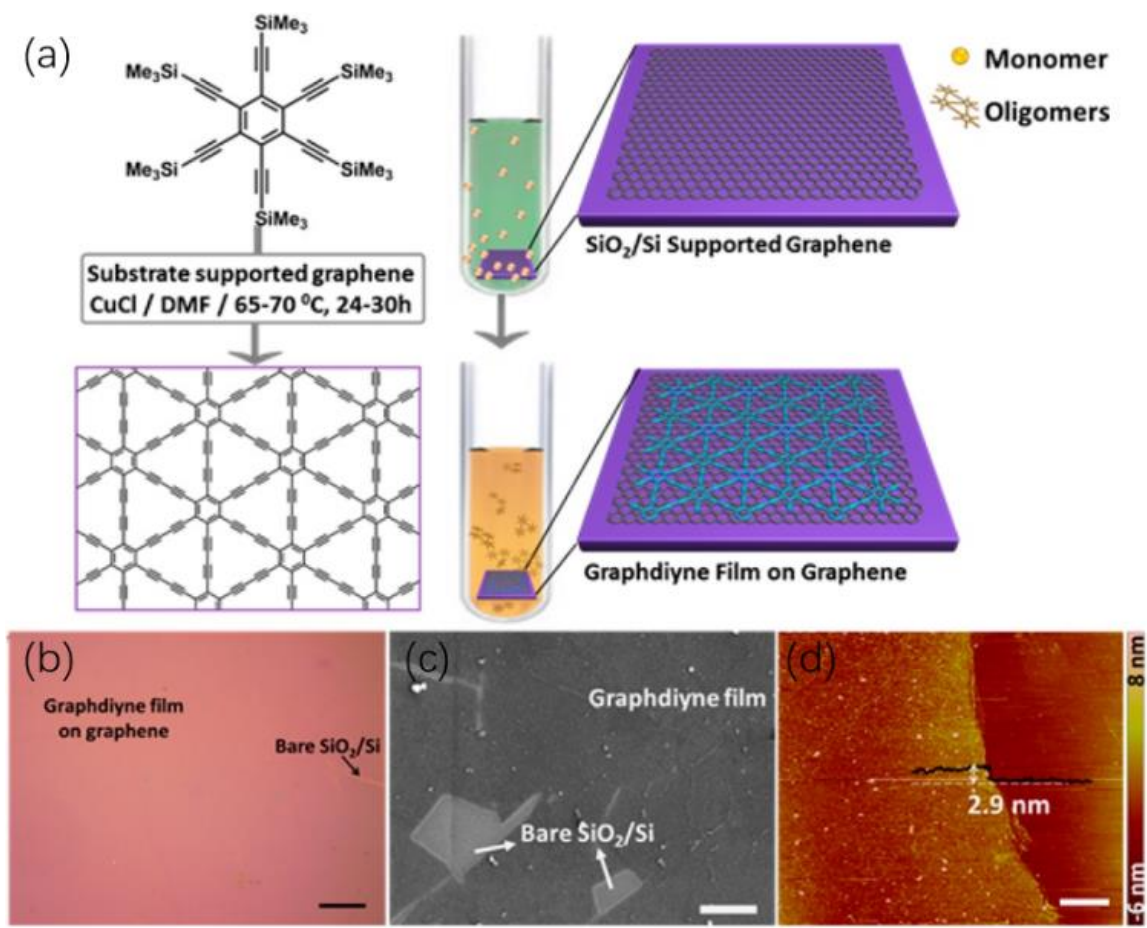

Figure 6. (a) Illustration of the preparation strategy of GDY film with graphene as substrate. (b) Optical, (c) AFM, and (d) SEM images of the as-prepared GDY film. Note that the region of bared SiO2/Si is marked by arrows. Reproduced with permission from ref. 121. Copyright 2018 American Chemical Society.

The GDY film was prepared by the coupling of nitrogen-containing substances as precursors [123], which could be bended in all directions to form a super-large transparent membrane.

3.1.2 Nanotube Arrays and Nanowires

Based on the previous work, Li et al. prepared the GDY array by using aluminum oxide plate [124], which was smooth and the wall thickness was about 40 nm. After annealing, the wall thickness was reduced by about 25 nm. If it was used as an emitter, it could be found that the pass field had a very low threshold value ranging from 8.83 to 4.20V/μm. The DTF method was used to calculate the electronic an[125]d structural properties of the new GDY array [126]. It's worth noting all the nanotubes under study exhibit semiconductor behavior. GDY powder was used to prepare GDY nanowires. The lengths of the prepared GDY nanowires ranged from 0.6 to 0.8nm and diameter ranged from 20 to 50nm, with a mobility of $7100V^{-1}s^{-1}$ and a conductivity of 1.9103 S $m^{-1}$. This feature indicated that GDY was used in optoelectronics and electronics fields [127].

GDY could be produced from copper oxide nanowires. Under the experimental environment of 180℃, a layered structure was formed on the copper oxide. At this time, GDY was formed in the outer layer and copper oxide was formed in the inner layer, as shown in Figure 7[128]. First, the copper hydroxide nanowires were prepared. Then, copper hydroxide was used as catalyst and matrix for the reaction. GDY was generated on the copper hydroxide nanowire. After annealing, the nanowires were transformed into copper oxide nanowires.

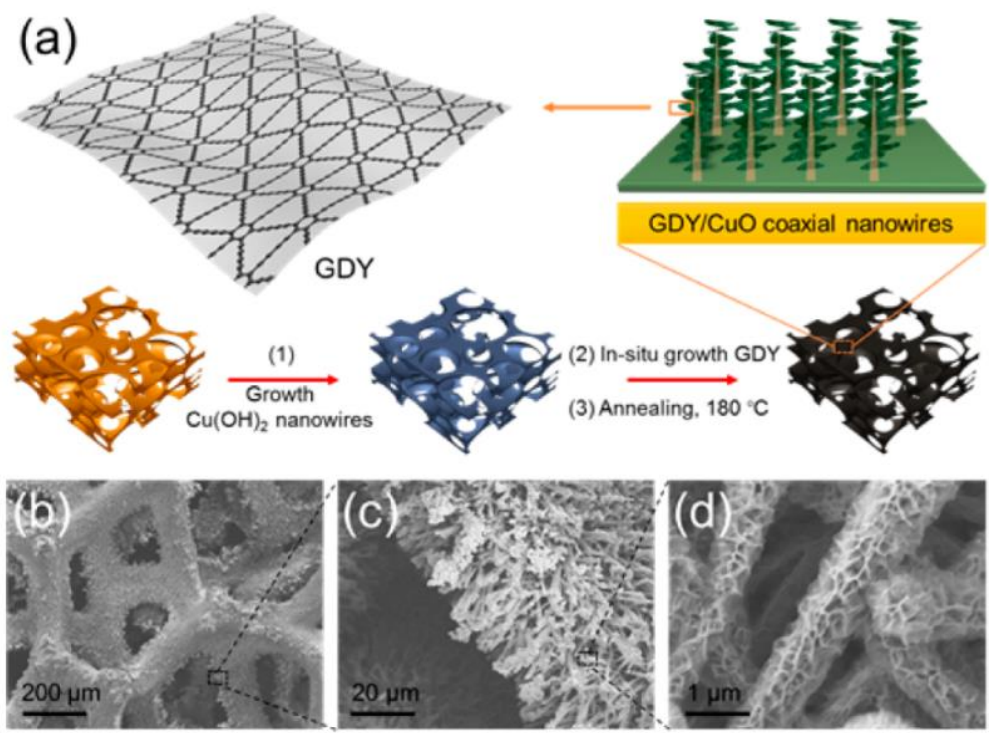

Figure 7 (a) Preparation process, (b)(c)(d)SEM images. Reproduced with permission from ref. [128]. Copyright 2017 American Chemical Society

3.1.3 Nanoballs

    Liu et al. proposed a new scheme based on coupling reaction to prepare GDY nanotube. As shown in Figure 8[129], copper plate was selected as the induction substrate and HEB was used as the precursor of the reaction. However, in the presence of a catalyst, copper was converted to copper ions, which acted as the catalytic point for the reaction. Firstly, the catalyst was used to rapidly generate GDY through coupling reaction. With the concentration of copper ions increasing, the newly generated GDY was attached to the prepared GDY and formed GDY nanowires. The GDY was transferred to the prepared substrate and the composition of prepared structure was displayed on the AFM image. GDY with high crystallinity could be obtained when the nano layer thickness was not equal. It was worth noting that the crystal gap was 0.466nm, which was consistent with the theoretical prediction. In later missions, the superhydrophobic foam of GDY was collected by using copper foam as the reactant [130]. In addition, GDY analogs could also be made using this method[131].

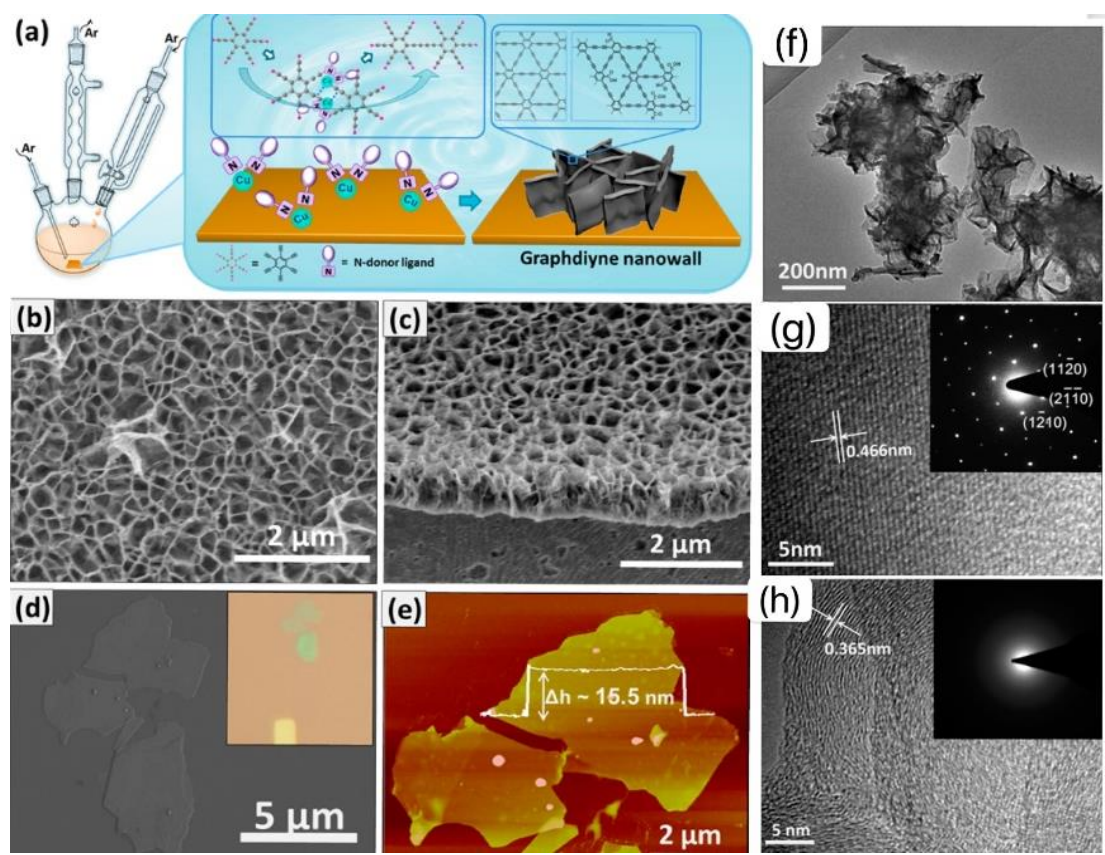

Figure 8. (a) Experimental schematic of preparing GDY nanowalls. SEM images of (b) the top and (c) the cross section for GDY nanoballs. (d) SEM and optical microscope image of the GDY sample exfoliated from GDY nanoballs. (e) AFM image of the exfoliated GDY sample with a thickness of ~15.5 nm. (f) TEM and (g, h) HRTEM images of GDY. The inset graphs show the SAED patterns. Reproduced with permission from ref. [129]. Copyright 2015 American Chemical Society.

Based on the preparation process of GDY nanowires, a scheme of preparing GDY nanowires using copper foil as catalyst was studied [132]. As shown in Figure 9, the role of copper foil was to ensure the concentration of copper ions. In this way, GDY could be produced on any base.

There is a preliminary understanding of GDY nanostructures, but the research findings on GDY are insufficient. With the in-depth study of the reaction process, it is expected that a large area of highly crystalline GDY will be synthesized in the future, and the basic properties of GDY will be further understood.

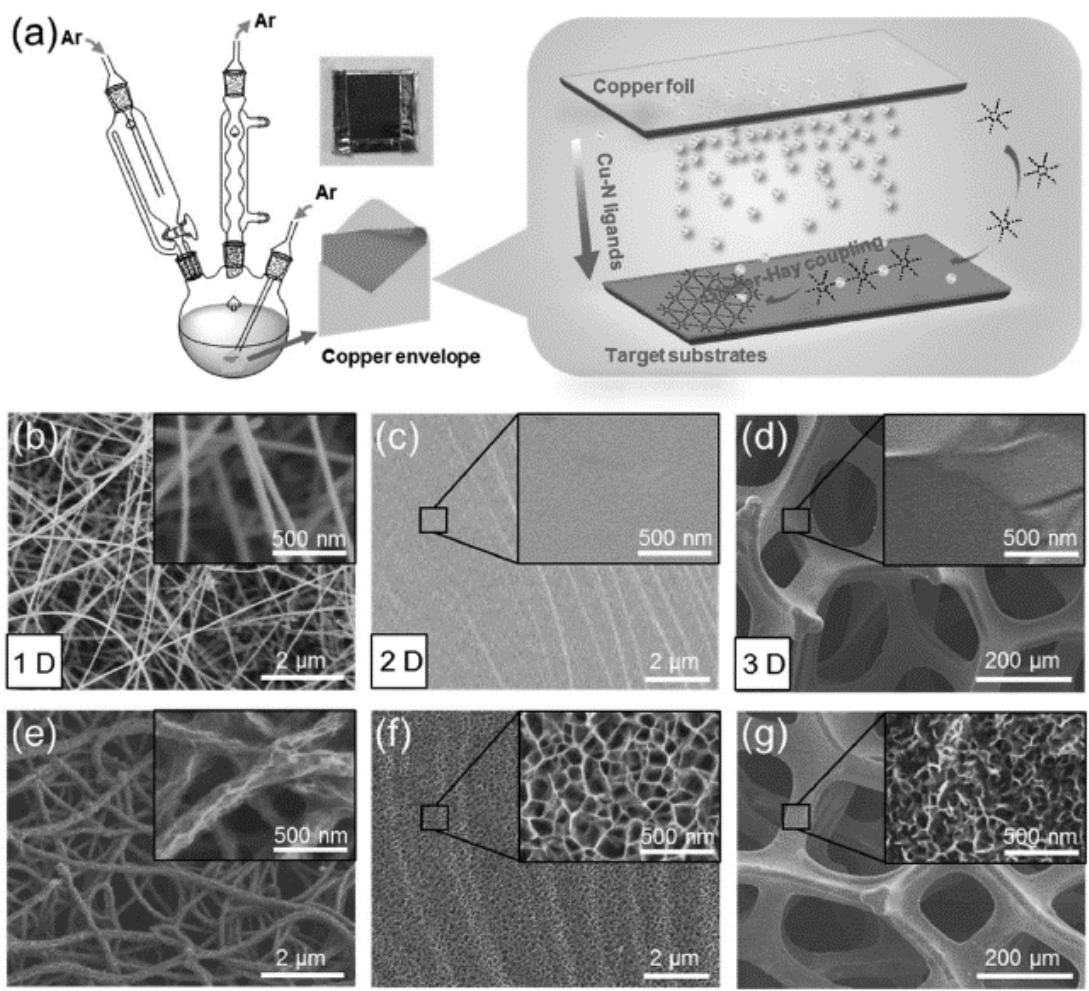

Figure 9. (a) Illustration of preparing GDY nanowalls following a method of the envelope copper catalysis. SEM images of the substrates: (b) silicon nanowires, (c) Au foil, and (d) GDY foam (GF) grown on Ni foam. SEM images of the GDY nanowalls grown on (e-g) the corresponding substrates. Reproduced with permission from ref. [132]. Copyright 2017 John Wiley and Sons.

3.1.4 Nanosheets

Sakamoto et al. synthesized GDY nanosheets using HEB as a precursor [133], as shown in Figure 10. Using copper acetate and pyridine as the catalyst for the coupling reaction, they were dispersed in the upper water layer. The precursor was dissolved in the lower layer of methylene chloride. GDY nanosheets was produced on the interface between the upper and lower layers. In the experiment, they chose to perform on the liquid/liquid or gas/liquid interface. In particular, the gas/liquid interface synthesized GDY nanosheets with single crystal features, which were regular hexagonal structures. The structure of the previously prepared GDY nanosheet had a side dimension of 1.5 μm and a thickness of 3 nm. After that, hexaethyltribenzene was selected as the reaction precursor, and a separate thin film could also be formed by the similar reaction described above [134]. The liquid/liquid interface could be used to synthesize nitrogen-substituted super films[135], and the thickness of the synthesized GDY film was 4 nm.

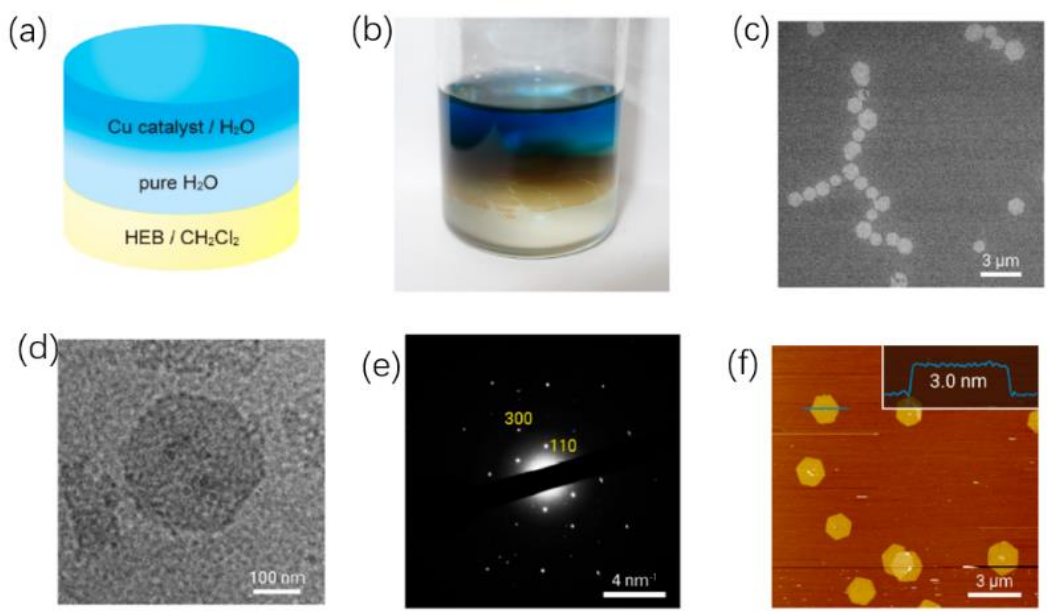

Figure 10. (a) Illustration and (b) an experiment setup of the synthetic procedure for GDY in the interface of liquid-liquid. Few-layer GDY (c) SEM micrograph. (d) TEM micrograph on an elastic carbon grid. (e) SAED pattern. (f) AFM image and the height analysis of the blue line region. Reproduced with permission from ref. [133]. Copyright 2017 American Chemical Society.

    H. Shang et al. had prepared ultra-thin nanosheets on independent copper nanowires. As shown in Figure 11, with CuNW as the reaction template [136], the crystal boundary of CuNW had a very high reactive, which provided a more efficient reaction site for the generation of GDY. The experimental GDY with a thickness of 3.75 nm had the same interlayer distance as the 0.365nm mentioned above. Through observation, copper nanoparticles could be found in the nanosheets, which showed that these particles had a strong force with the GDY nanosheets. GDY nanotubes were produced when placed CuNW in the mixed solution of HCL and FeCL3.

    This method was a new method for ultrathin GDY nanosheets. The large-scale production of ultra-thin GDY nanostructures can be realized. The crystal boundary of copper substrate was highly reactive to the growth of GDY, so the nanostructure and quality of copper substrate are strictly required.

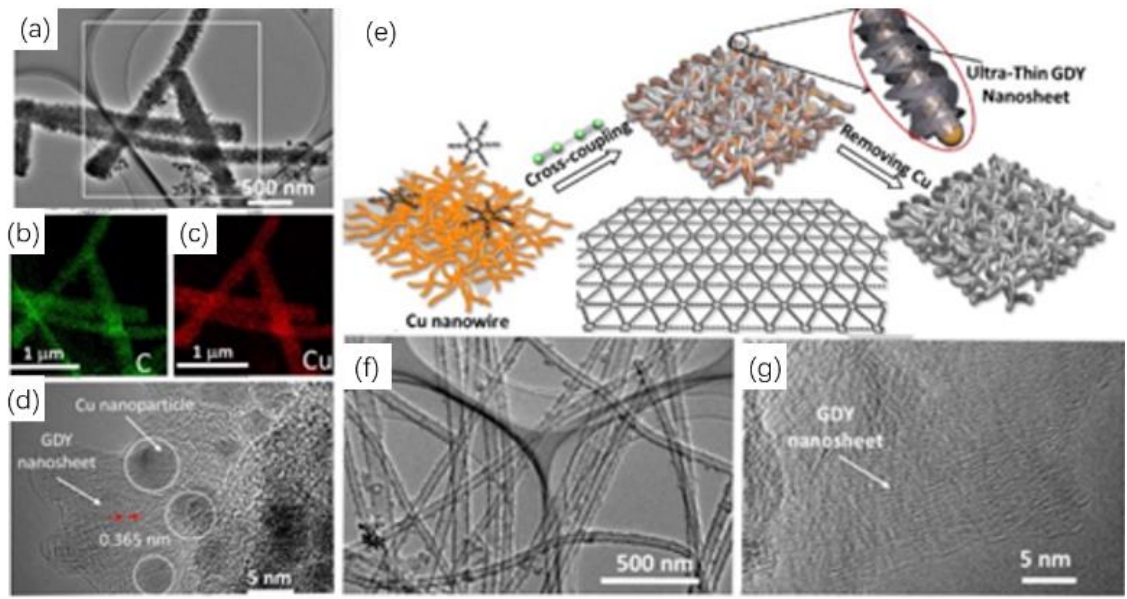

Figure 11 (a) TEM and (b)(c) energy-dispersive spectrometry (EDS) of GDY nanosheets on CuNW. (d) with Cu. (e) Schematic illustration of preparing GDY on CuNW. (g) GDY nanotube TEM image and (f) without Cu. GDY nanosheets HRTEM images: Reproduced with permission from ref. [136]. Copyright 2018 John Wiley and Sons.

3.1.5 Ordered Stripe Arrays

GDY had great elasticity and electronic properties, therefore it could be made into a stretchable sensor. In addition, in the process of producing GDY with copper foil, the graphicalization of GDY could be realized if the silicon templates with different shapes were determined in advance，as shown in Figure 12[137]. The template with strong hydrophilicity could make the reactant diffuse to the surface of the substrate continuously. In this way, the shape of GDY could be precisely controlled in electronic devices.

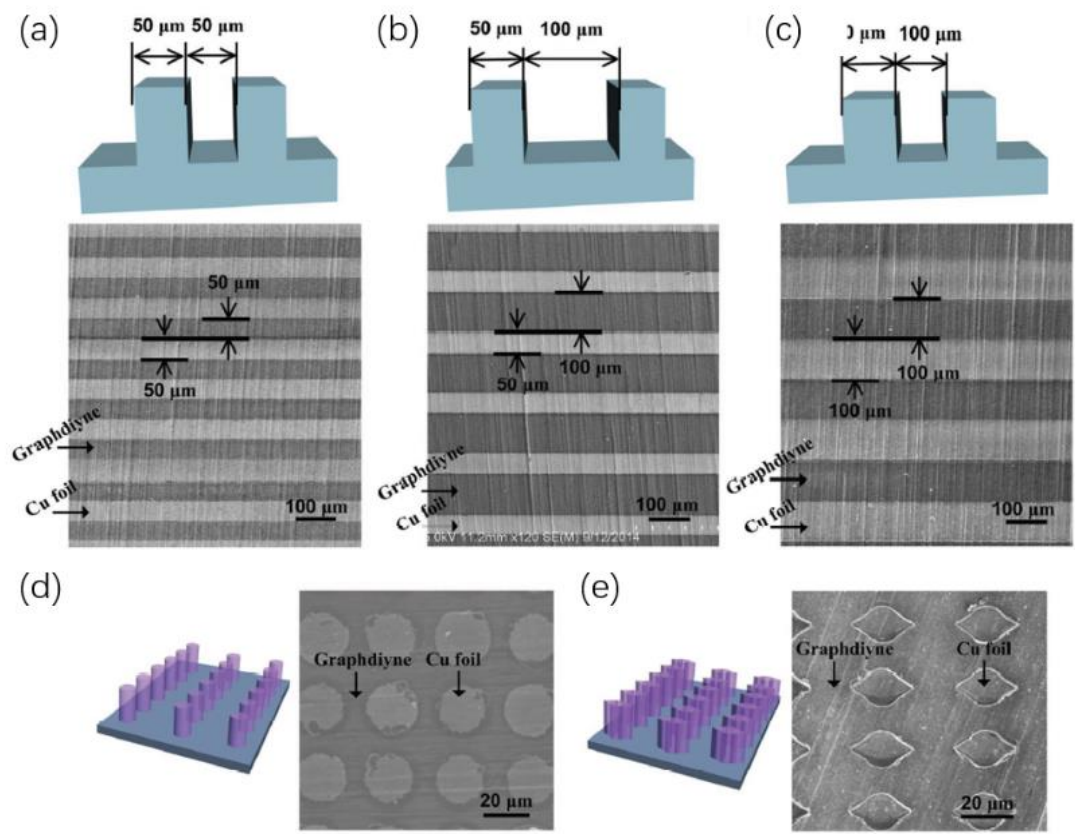

Figure 12. Illustration of silicon templates and corresponding SEM images of GDY patterns on silicon templates: (a)(b)(c) predefined groove with different size, (d) circle, and (e) spindle. Reproduced with permission from ref. [137]. Copyright 2017 John Wiley and Sons.

3.1.6 GDY 3D Framework

GDY had been synthesized, but it was still difficult to produce GDY in large quantities. In recent years, Li et al. had been found that the mass production of GDY with different nanostructures could be achieved by explosive method[138]. As shown in Figure 13, the method was to heat the HEB under the condition of 120℃ with non-metal catalyst. If the HEB was put into the air that had been heated, the material reacted violently and the productive rate of the GDY sample reached 98%. This method could also adjust the structure of GDY, N-GDY nitrogen concentration and structure were adjusted to display excellent controllability[139].

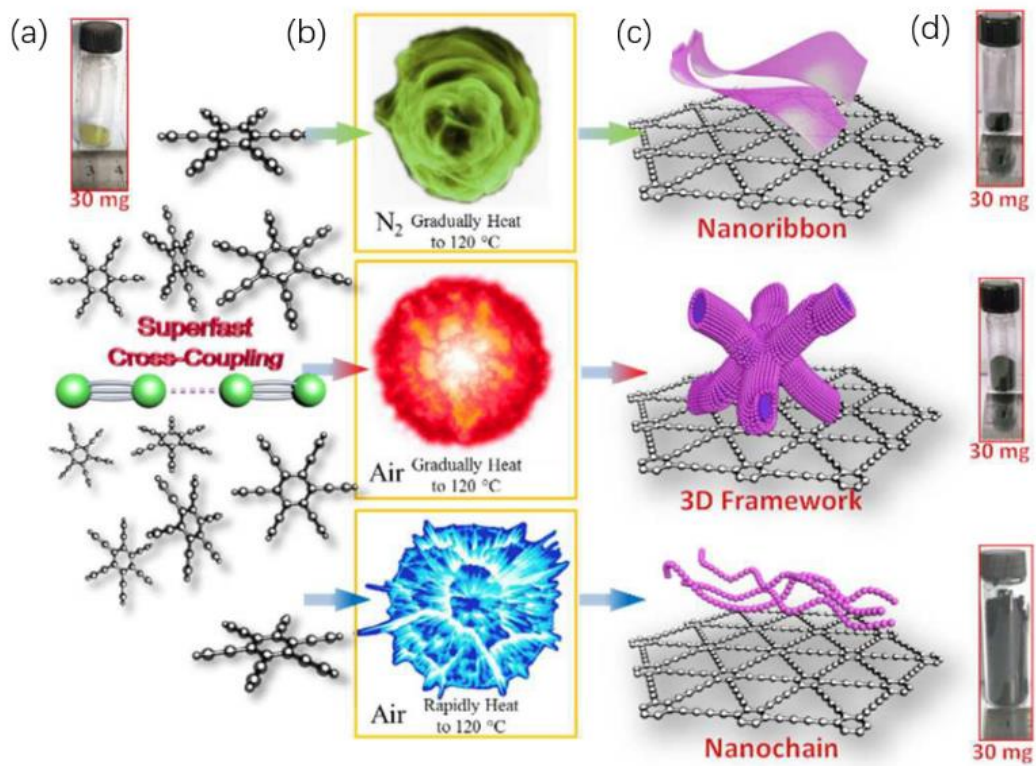

Figure 13 Illustrations of the preparation processes. (a) Photo of HEB before the reaction; (b) reaction under three treatments; (c) GDY morphologies; and (d) sample photos showing the volume change after the reaction. Reproduced with permission from ref. [138]. Copyright 2017 Royal Society of Chemistry

Zhang et al. also proposed other ways to make novel GDY structure, and diatomite was selected as the substrate. As shown in Figure 14[140], diatomite was abundant and highly absorbent. Therefore, copper nanomaterial was placed in the interstitiums of diatomite. Copper as the catalyst for the reaction was adsorbed on the template to prepare GDY. The three-dimensional GDY structure could be reinforced by connecting hollow GDY columns. Diatomite and nanometer copper were etched to form three-dimensional GDY.

Noble metals have a wide range of applications in electrochemical catalysis and organic synthesis. It has a larger surface area and more active sites, hence it has a higher catalytic efficiency. However, it is easy to polymerize, difficulty in recycling, and high cost limit its application.

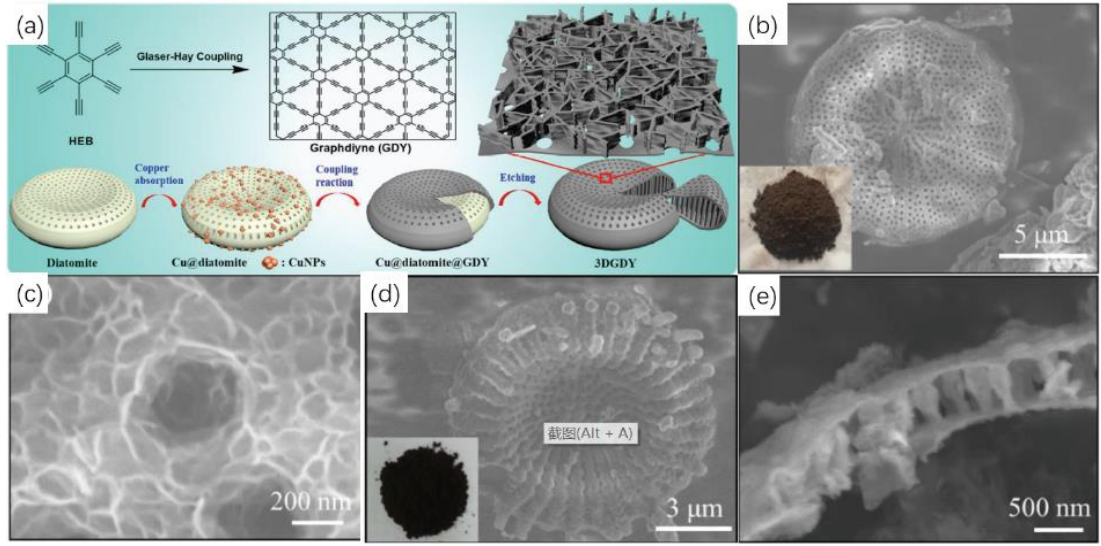

Figure 14. (a) Preparation processes of 3D GDY. SEM images of GDY grown on diatomite with Cu nanoparticles being absorbed: (b) zoomed-out and (c) zoomed-in images. SEM images of freestanding 3D GDY without diatomite template: (d) zoomed-out and (e) zoomed-in images. Reproduced with permission from ref. [140]. Copyright 2018 John Wiley and Sons.

**3.2 Mechanical properties**

The mechanical properties of carbon materials are very representative. GR and CNT are good examples in these materials. Graphynes have an unusual pore structure and it is worth further study. Wang et al. explored and applied a different molecular dynamics method. Wang concluded that the acetylenic linkages is the key. The presence or absence of it has an important influence on the mechanical properties of graphynes, the results are shown in Table 2[141]. In the case of increasing percentage of acetylenic linkages. $\gamma$-, $\beta$- and $\alpha$-graphynes are in order from high to low. That is, the order of fracture stress and Young's modulus. 6,6,12-graphyne is quite special in terms of its mechanical properties. This special manifestation is that the Young's modulus is different in different directions. For example, it is 0.445 TPa in the horizontal direction, but 0.35 TPa in the vertical direction. From another perspective, the acetylenic linkages have an effect on the fracture strain, it is indirectly affected by flexibility, which can be seen from Table 2.

Table 2 Fracture stresses, strains and Young's modulus of graphynes[141].

| Model | Atom density (atoms per nm$^2$) | Stress(Gpa) | | Difference in stresses (%) | Strain | | Difference in strains (%) | Young's modulus (TPa) | |
| --- | --- | --- | --- | --- | --- | --- | --- | --- | --- |
| | | $x$ | $y$ | | $x$ | $y$ | | $x$ | $y$ |
| $\alpha$-Graphyne | 18.92 | 36.36 | 32.48 | 10.69 | 0.178 | 0.156 | 12.37 | 0.12 | 0.119 |
| $\beta$-Graphyne | 23.13 | 46.26 | 38.06 | 17.72 | 0.162 | 0.130 | 19.54 | 0.261 | 0.26 |
| 6,6,12-Graphyne | 28.02 | 61.62 | 39.06 | 36.61 | 0.147 | 0.116 | 21.54 | 0.445 | 0.35 |
| $\gamma$-Graphyne | 29.61 | 63.17 | 49.78 | 21.20 | 0.148 | 0.112 | 24.09 | 0.505 | 0.508 |
| Graphyne | 39.95 | 125.2 | 103.6 | 17.27 | 0.191 | 0.134 | 29.93 | 0.995 | 0.996 |

Li et al. used the first-principles method to study graphynes. The mechanical properties were further calculated[93]. Li calculated their bandgap changes, and they were the first in this regard.

The premise is that the application of different tensile stress. Wang et al. have reached similar conclusions. The more acetylenic linkages, the more carbon-carbon single bonds will increase. This leads to a reduction in the in-plane intensity of graphyne. There is a mathematical relationship between the intensity of $C$ and $n$. Where $C$ is the stiffness and $n$ is the number of acetylenic linkages. And GR is greater than all of them. From the perspective of the band structure, they concluded that the strain pattern is related to the change in bandgap, as shown in Figure 15 and Figure 16.

It can be seen from Figure 15(b) and Figure 15(d) that M is the direct gap point, S point too. The change in strain pattern corresponds to different bandgap changes. Under homogeneous tensile stretching it increases. But in the case of uniaxial tensile stretching, the opposite is true. However, the direct gap point has changedin in graphyne-4, it is always the $\Gamma$ point. The shift of energy state is the main reason for the change of band structure. The above process has studied the mechanical properties of graphynes.

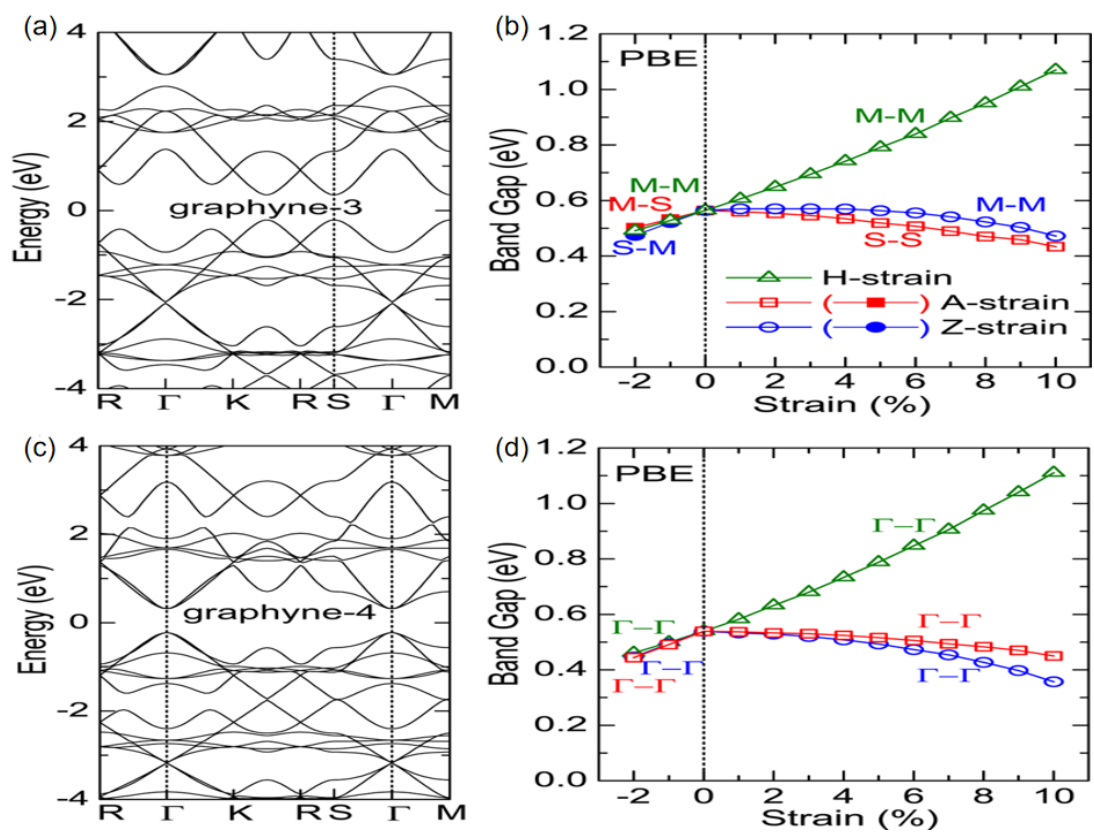

Figure 15. (a) Band structure of graphyne-3 in the absence of stress. (b) Different stresses are applied and the band structure changes with strain. GGA-PBE functional was applied to calculate the results. (c) and (d) have the same results as (a) and (b), but the object of calculation is graphyne-4. Reprinted with permission from ref.[93]. Copyright 2013 American Chemical Society

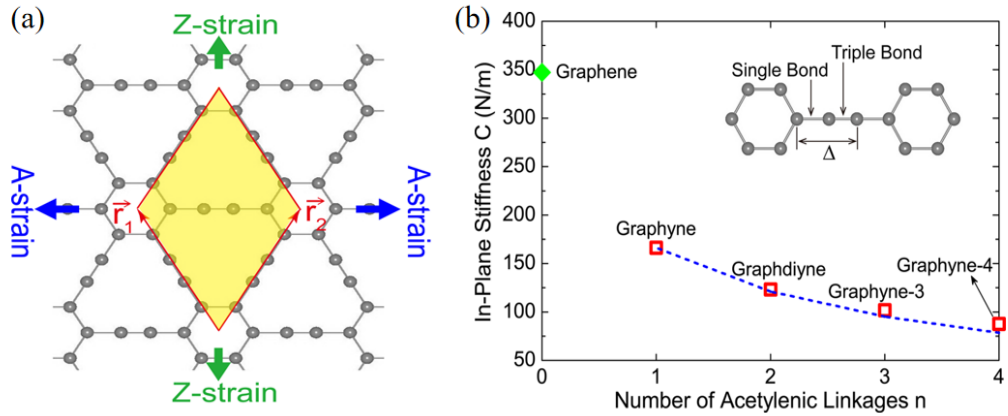

Figure 16. Geometrical structure of GDY. The stress direction is indicated. Relationship between in-plane stiffness, $C$, and number of acetylenic linkages $n$. The maximum one is the $C$ value of graphene. Reprinted with permission from ref.93. Copyright 2013 American Chemical Society

**3.3 Electronic properties**

3.3.1 Electronic properties of single-layer graphynes.

GR is a semiconductor material with zero bandgap, it has attracted a lot of attention. GDY is a chemically synthesized carbon allotrope material. Therefore, there is more and more research on it. In theory, graphyne 's electrical properties may be better than GR[104]. Next, the electronic properties of GDY are studied, such as the narrow band gap properties. The narrow band gap of single-layer GDY is 0.44-1.47 eV. And the main reason why the value is within a certain range is that the calculation method used is different[45,142]. Figure 16 shows the corresponding situation of GDY. Obviously, the bandgap value at point $\Gamma$ can be seen[61]. Mobility is a crucial parameter and it can be used to measure the electrical properties of semiconductor materials. And it also plays a decisive role in the application of semiconductor electronics.

Table 3 gives some data of several single-layer graphynes and GR. The hexagonal structure of 6,6,12-graphyne deserves attention. But it does not have the symmetry of polygons. In addition, the structure also shows anisotropy. When the selected direction is fixed, the mobility of GR on it is much smaller than that of graphyne. It shows that graphyne has better electronic properties.

Table 3 Intrinsic hole and electron mobilities (300 K) and bandgap of single-layer graphyne material obtained by different calculation methods

| Carbon allotropes | Hole mobility $\mu_h$ ($10^4$ cm$^2$ V$^{-1}$s$^{-1}$) | Electron mobility $\mu^e$ ($10^4$ cm$^2$ V$^{-1}$s$^{-1}$) | Bandgap (eV) |
|---|---|---|---|
| $\alpha$-Graphyne | 3.316,[143a] 2.960,[143b] 0.96[144] | 3.327,[143a] 2.716,[143b] 1.03[144] | 0[143] |
| $\beta$-Graphyne | 1.076,[143a] 0.856[143b] | 0.892,[143a] 0.798[143b] | 0[143] |
| $\gamma$-Graphyne | 0.35[144] | 1.62[144] | 0.454,[144] 0.47,[50] 0.471,[145] 0.48,[146] 0.52,[58] 1.32,[142] 2.23[50] |
| 6,6,12-Graphyne | 42.92,[143a] 12.29[143b] | 54.10,[143a] 24.48[143b] | 0[143] |
| γ-Graphdiyne | 1.97,[143a] 1.91[143b] | 20.81,[143a] 17.22[143b] | 0.44-1.47[45,50,58,61,142,146-149] |
| Graphyne | 32.17,[143a] 35.12,[143b] 23.49[144] | 33.89,[143a] 32.02[143b], 20.05[144] | 0[150] |

[a] Zigzag direction. [b] Armchair direction.

3.3.2 Electronic properties of GDY with different layers and stacking models.

The difference in layers will affect the electronic properties. The same is true for the stacking method. Lu et al. studied this situation of GDY[149]. Figure 17 shows the most stable accumulation pattern of GDY, including bilayer and trilayer GDY. These two materials also have metallic properties. Figure 18 shows the new stacking model of GDY proposed by Hiroshi [133] and Zhang et al[151]. In the experiment, they used the ABC stack to measure the photoluminescence (PL) spectrum of the three-layer GDY. They also observed an emission peak with a photon energy of 1.79 eV, which is very close to the predicted band gap of the three-layer GDY. In addition, they predicted the band structure and band gap of single-layer, double-layer, and triple-layer GDY, and concluded that the band gap of GDY decreases as the number of layers increases.

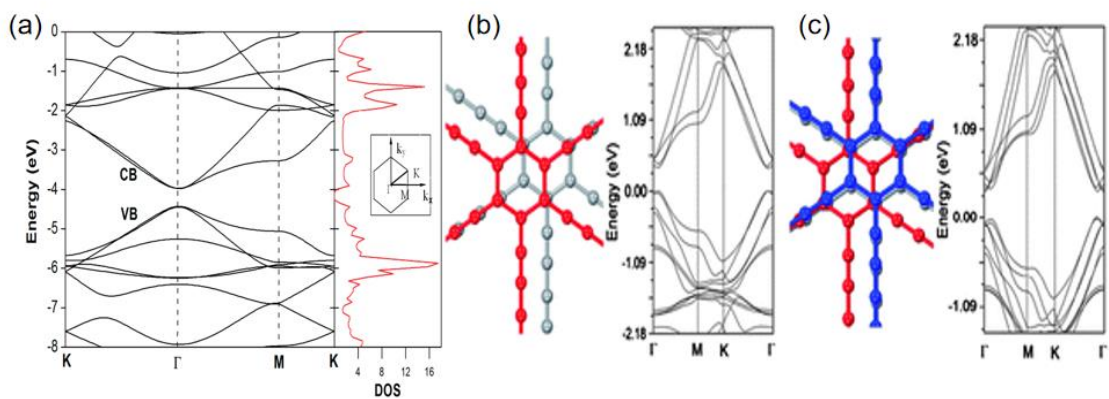

Figure 17. (a) Band structure of single-layer GDY; reprinted with permission from ref.61. Copyright 2011 American Chemical Society. (b) Most stable conformation, AB (b1), of the bilayer GDY and its energy band structure. (c) Most stable conformation, ABA, of the trilayer GDY (g1) and its band structure. Reprinted with permission from ref.[149]. Copyright 2012 The Royal Society of Chemistry.

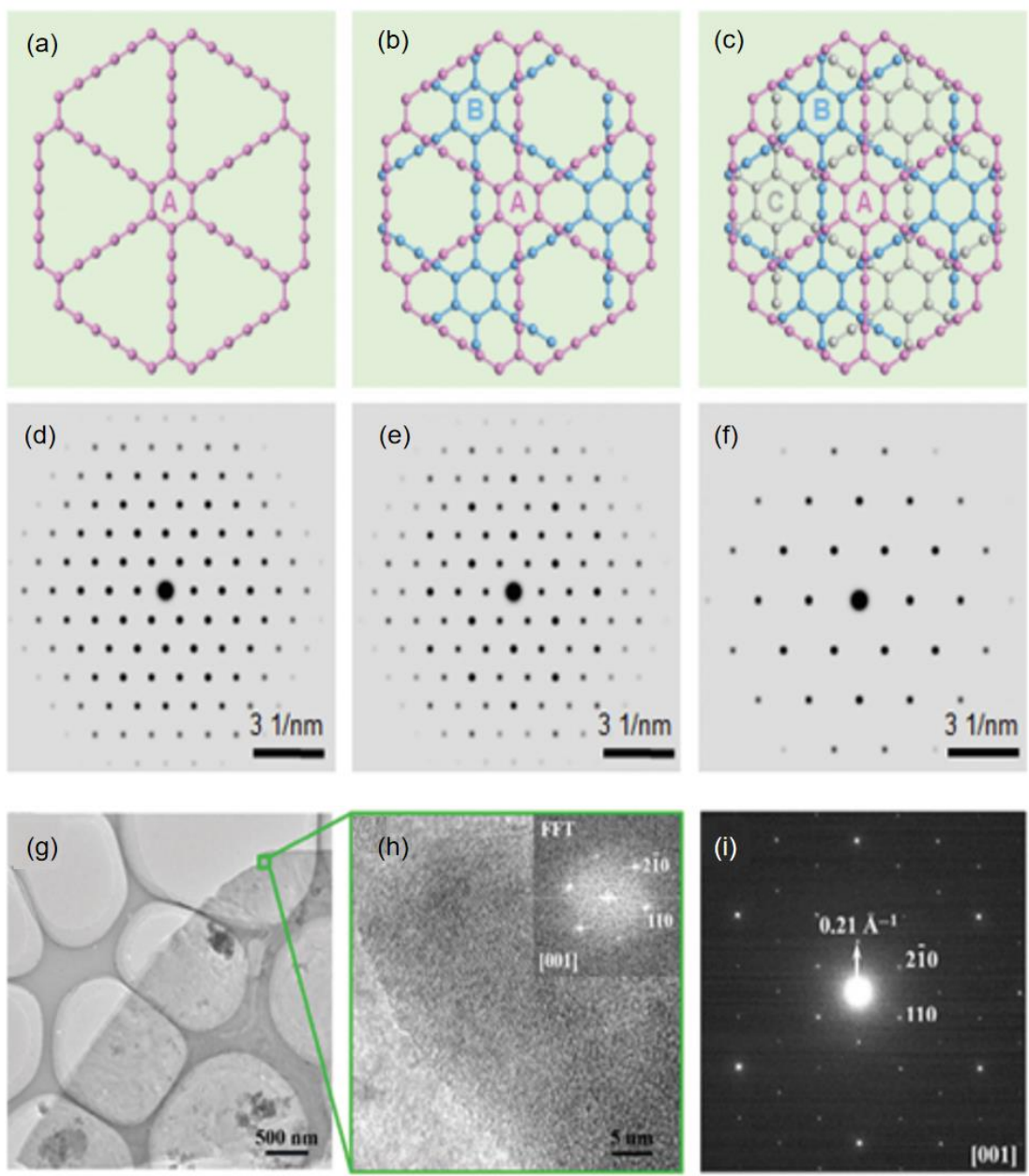

Figure 18. SAED pattern and HRTEM image of few-layer GDY sheets. (a)(b)(c)(d)(e)(f) Considered stacking structures and corresponding TEM/SAED simulation patterns from the top view. AA stacking (a and d), AB stacking (b)(e), and ABC stacking (c)(f). (g) Low-magnification TEM image of six-layer GDY nanosheets. (h) HRTEM image of the circled area in (g). Inset is the fast Fourier transform (FFT) pattern of the HRTEM image. (i) SAED pattern of GDY nanosheets. The zone axis is [001]. Reprinted with permission from ref.[152]. Copyright 2018 The Royal Society of Chemistry.

### 3.4 Optical properties

3.4.1 Optical absorption.

When it comes to optical properties, it must be said that optical absorption. Lu et al. studied this property of GDY and calculated its spectrum, including the ultraviolet-visible-near-infrared spectrum[148]. Li transferred the as-grown GDY [151] to the substrate. This substrate is made of quartz glass. Figure 19(a) shows the corresponding absorption spectrum. Figure 19(b) shows the absorbance after the red line is subtracted. It is then compared with the theoretical results. A result is calculated by Bethe-Salpeter equation. Then compare GW+RPA and experimental results with them respectively. The conclusion is that the experimental results are more consistent. Different peaks will appear in different transitions. Transitions include those around band gaps and those around Van Hove singularities. The first peak is caused by the first transition, while the other peaks are caused by the second transition. Liu et al. got almost the UV-vis spectrum[153]. But Figure 19(c) shows that it has a significant bathochromic shift. The main reason for this phenomenon is electronic enhancement.

3.4.2 Raman spectroscopy of GDY.

Raman scattering is a key piece of information. It can reflect the structure of GDY. And this information is more important for Raman-active diyne linkers. Liu et al. studied Raman spectrum[154]. GDY has a special structure, it has many atoms and acetylenic linkages, so it has many vibration modes, Figure 19(d) and (e) show several of them. Peaks Y and Y' in Figure 19(d) are more representative. They come from two different modes of alkyne triple bond. Due to the stretching of the aromatic bonds, GDY has a G peak. But unlike GR, the G peak here has a bathochromic shift.

Liu et al. studied the representative spectrum in detail. But the experimental results were unsatisfactory. The reported Raman spectrum [60] is different from the prediction. As shown in Figure 19(f), the main difference is resolved and bands. Zhang et al. made a new type of GDY film. Among them, GR is used as a substrate[151]. Figure 19(g) shows the results of their experiment. They got a perfect Raman spectrum. And it can be well matched with the predicted spectral characteristics.

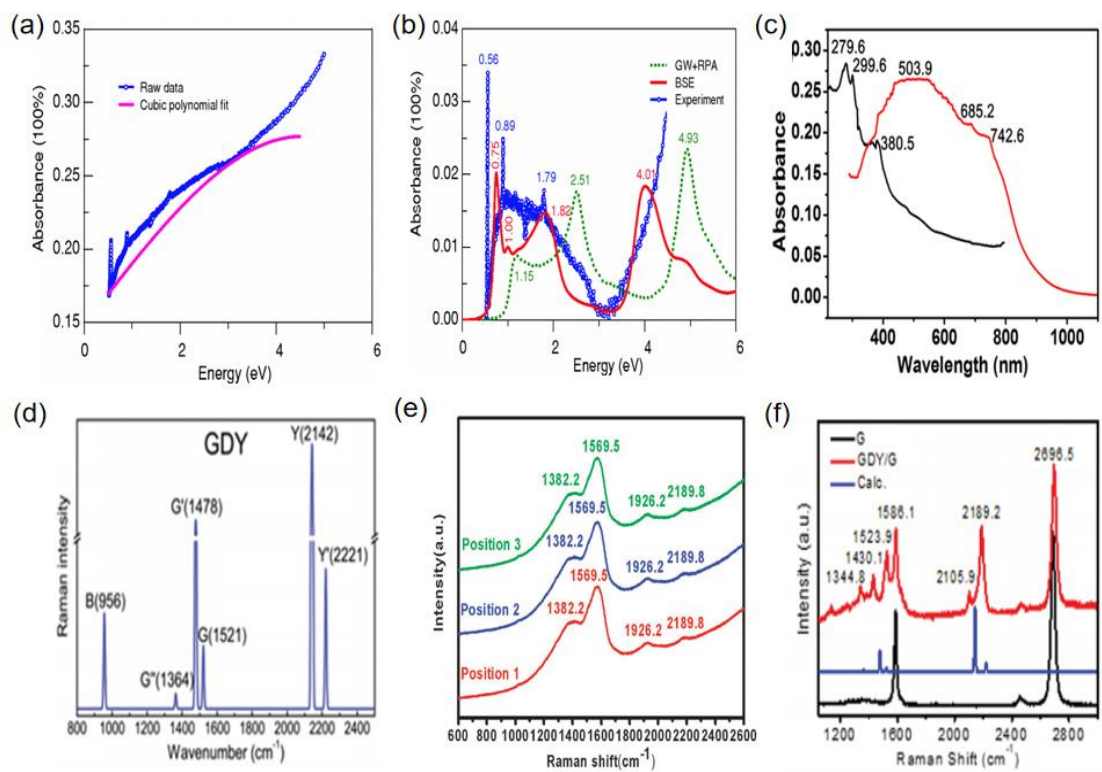

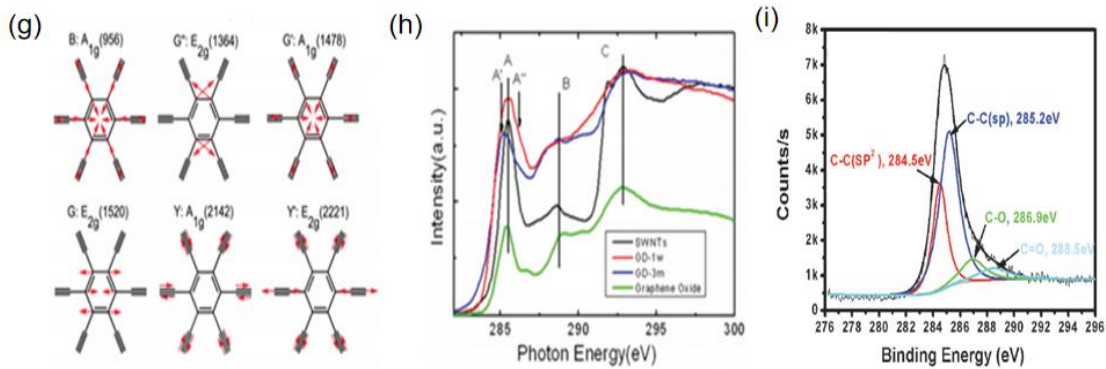

Figure 19. Optical properties of GDY. UV-vis-NIR spectra of GDY (a) before and (b) after subtracting the background. Reprinted with permission from ref.[148]. Copyright 2011 American Physical Society. (c) UV-vis spectra of HEB monomers (black line) and GDY nanowalls (red line). Reprinted with permission from ref.[153]. Copyright 2015 American Chemical Society. (d) Predicted Raman spectrum, whose peaks are with 10 cm1 Gaussian smearing. Reprinted with permission from ref.[154]. Copyright 2016 American Chemical Society. (e) Raman spectra of the as-grown GDY at three different locations on the Cu foil. Reprinted with permission from ref.60. Copyright 2010 The Royal Society of Chemistry. (f) Raman spectra of GR (black line), GDY/GR (red line), and predicted Raman spectrum of GDY (blue line). The Raman spectra were collected using 514.5 nm excitation laser. Reprinted with permission from ref.[151]. Copyright 2018 American Association for the Advancement of Science. (g) Atomic motions of intense Raman-active modes. The red arrows show the motion directions of the main contributors. Reprinted with permission from ref.60. Copyright 2016 American Chemical Society. (h) XAS of GDY. Reprinted with permission from ref.[155]. Copyright 2013 American Chemical Society. (i) High-resolution asymmetric C 1s XPS spectrum of the as-grown GDY. Reprinted with permission from ref.[60]. Copyright 2010 The Royal Society of Chemistry.

3.4.3 X-ray absorption (XAS).

Figure 19(h) shows the different results of Wang et al characterized by XAS[155]. They studied the K-edge spectrum of carbon. Comparing the spectrum of CNT with GDY, it can be seen that it is different. GDY's A peak is called A". Wang et al. continued to study the functional groups possessed by GDY and analyzed their absorption spectrum. The stability of GDY is a key point. It can be judged by putting it in the air. The blue line in Figure 19(h) gives the required XAS spectrum. The GDY was exposed to the air for up to three months. Figure 19(i) shows the XPS spectrum of GDY. It can be seen that it is different from CNT and GR[60].

**3.5 Magnetic properties**

It is well known that spintronics is widely used. There is more and more research on the inherent magnetic properties of low-dimensional carbon. Many studies have proved that the main causes of magnetism are vacancies and sp-type defects in two-dimensional carbon materials. Because they destroy the delocalized pi electronic system and prevent the aggregation of sp-type functional groups.

GDY's unique structure is worth studying. It can effectively avoid functional group clustering. Tang et al. studied the intrinsic magnetism of GDY[156]. It was found to have spin-half paramagnetism. Then they studied the corresponding properties of the prepared GDY. They finally found the magnetic source. The hydroxyl groups on GDY is the key to magnetic properties.

Because it has a relatively high migrating barrier energy. In this way, they can effectively prevent them from gathering. The paramagnetism of GDY can be changed. GDY can be annealed in an ammonia atmosphere. It can be found that the saturation moment increases at 2 K [156].

**3.6 Properties of functional GDY**

There are many 3d transition metal (TM) atoms, for example, V, Cr, Mn, etc. They can be adsorbed on GDY. At the same time, it will produce electrons and magnetism. Sun et al. conducted research on this[146]. They came to a conclusion that when the onsite Coulomb energy value changes, it will affect TM-GY/GDY. The effect is to change their electronic structure. The adsorbed atoms can adjust the electronic structure. They can also induce excellent magnetism.

The corresponding properties are shown in Figure 20. Two conclusions can be drawn. The first conclusion is that the adsorption of certain atoms does not cause magnetism, as shown in Figure 20(a). Ni-GY and Ni-GDY are good example. The second conclusion is that the adsorption of atoms can adjust the band structure, as shown in Figure 20(b). Co-GY, etc. behave as spin-polarized half-semiconductors. Fe-GDY, etc. behave as metals. Ni-GY / GDY has nonmagnetic semiconductor properties.

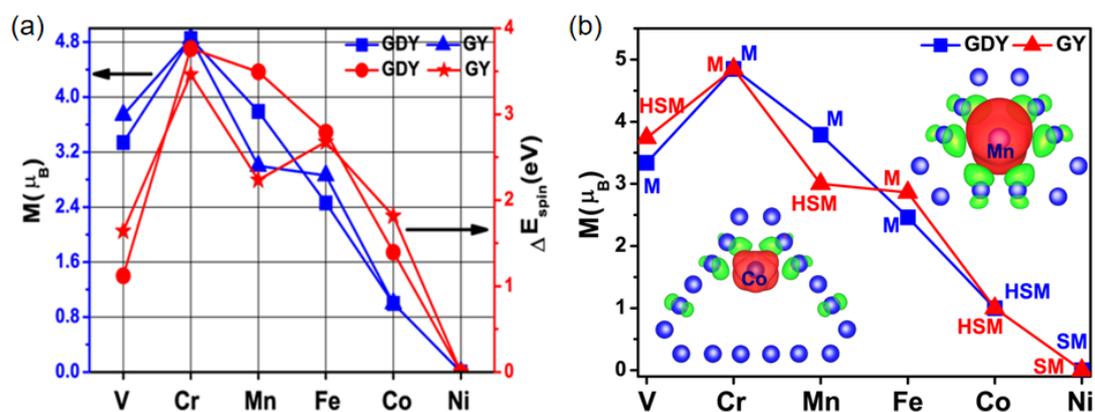

Figure 20. (a)Magnetic moment(M) and spin-polarized energy, DEspin, values of single-TM-atom-absorbed GY and GDY. (b) Magnetic properties of singleTM-atom-absorbed GY and GDY. "M," "HSM," and "SM" denote metal, half-semiconductor, and semiconductor, respectively. Reprinted with permission from ref.[146]. Copyright 2012 American Chemical Society.

# 4. Application of GDY in catalysis

GDY is a two-dimensional planar network structure of all carbon molecules, which is formed by connecting benzene rings with 1,3-diynebonds. It has relatively many carbon chemical bonds, a great conjugated system, and prominent chemical properties. In addition, GDY has great potential in the field of electronics and energy [35,47,157-168]. In summary, GDR materials are excellent catalyst, energy and sensor materials. Many researchers have used GDR materials as catalysts for gas separation and water purification, and at the same time achieved good experimental results.

Carbon-based catalysts have some of the same important characteristics, such as larger contact area, stable gaps and structure. The same is true for the GDY catalyst, due to its two-dimensional conjugated system, the GDY catalyst can increase the reaction area. When multiple catalysts work together, GDY catalyst can also improve the stable working environment.

Recently, a large number of GDY catalysts have been applied to the market and have shown great potential.

## 4.1 Photocatalysts

The photocatalyst has a strong redox capacity under light, so as to achieve the purpose of purifying pollutants and other catalytic purposes. In addition, the energy required for photocatalysis is relatively low, and there is no secondary pollution. These characteristics make photocatalysis technology widely recognized and applied. As a material with strong adhesion, non-toxic and stable chemical properties, $TiO_2$ is widely used in photocatalysis. $TiO_2$ has a strong barrier to ultraviolet rays with a wavelength of 190-400nm. GDY is a two-dimensional planar semiconductor which can improve the photocatalytic performance with $TiO_2$. At present, many researchers combine GDY catalyst with $TiO_2$ for photocatalysis [64,169,170]. Hydrothermal reaction can be used to decompose insoluble materials or pre-treat samples. GDY is added to the hydrothermal reactor for converting part of the diynyl group into a two-dimensional conjugated structure by hydrothermal reaction. The activity of $TiO_2$ catalyst can be improved by a variety of two-dimensional carbon materials, such as GDY, graphene and carbon nanotubes, as shown in Figure 21[169]. In the visible and ultraviolet bands, the photocatalytic activity of $TiO_2$ combined with GDY is higher than that of $TiO_2$ combined with other materials. In this meanwhile, the combination of $TiO_2$ and GDY greatly increase the adsorption of the catalyst in the visible and ultraviolet bands compared to other materials. The experimental results show that adjusting the $TiO_2$ content in the catalyst can further optimize the catalytic performance. The optimized value can be increased to 0.6wt%.

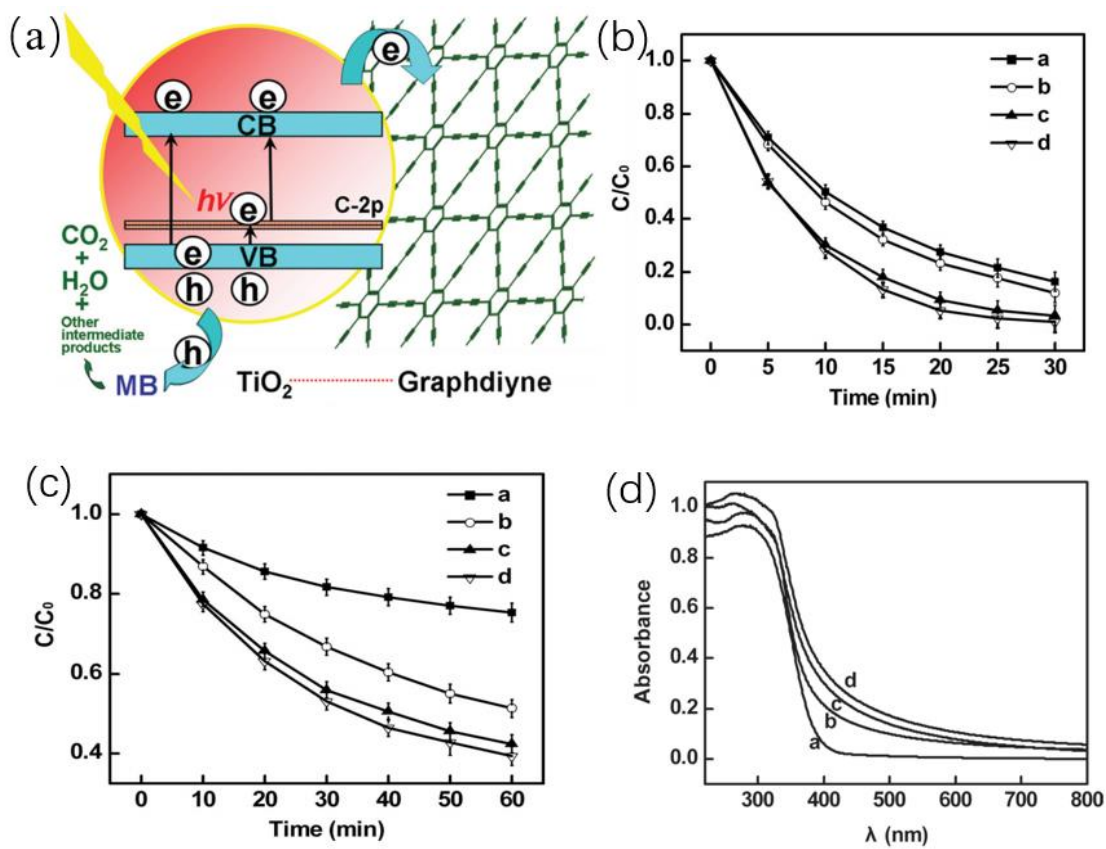

Figure 21(a). $TiO_2$ and GDY composite diagram. (b). Relationship between photocatalytic and time under ultraviolet light. (c). Relationship between photocatalytic degradation and time under visible

light. (d). Diffuse reflectance spectra of nanocomposites under ultraviolet and visible light. Reproduced with permission from ref. [169]. Copyright 2012 John Wiley and Sons.

The first principle is used to study different $TiO_2$ composite materials [47][64]. A comparison between the $TiO_2$ graphene composite materials and the $TiO_2$ GDY materials shown that the $TiO_2$ GDY material has higher oxidation capacity and service life, and its performance is better than the $TiO_2$ graphene composite. The valence band position of the material composed of $TiO_2$ and GDY is lower than that of the pure $TiO_2$ or $TiO_2$ graphene composites, as shown in Figure 22[64]. Therefore, $TiO_2$ GDY is used as a photocatalyst, and its catalytic rate is 1.63 times that of pure $TiO_2$ catalyst. $TiO_2$ GDY materials as photocatalyst has a huge market prospect.

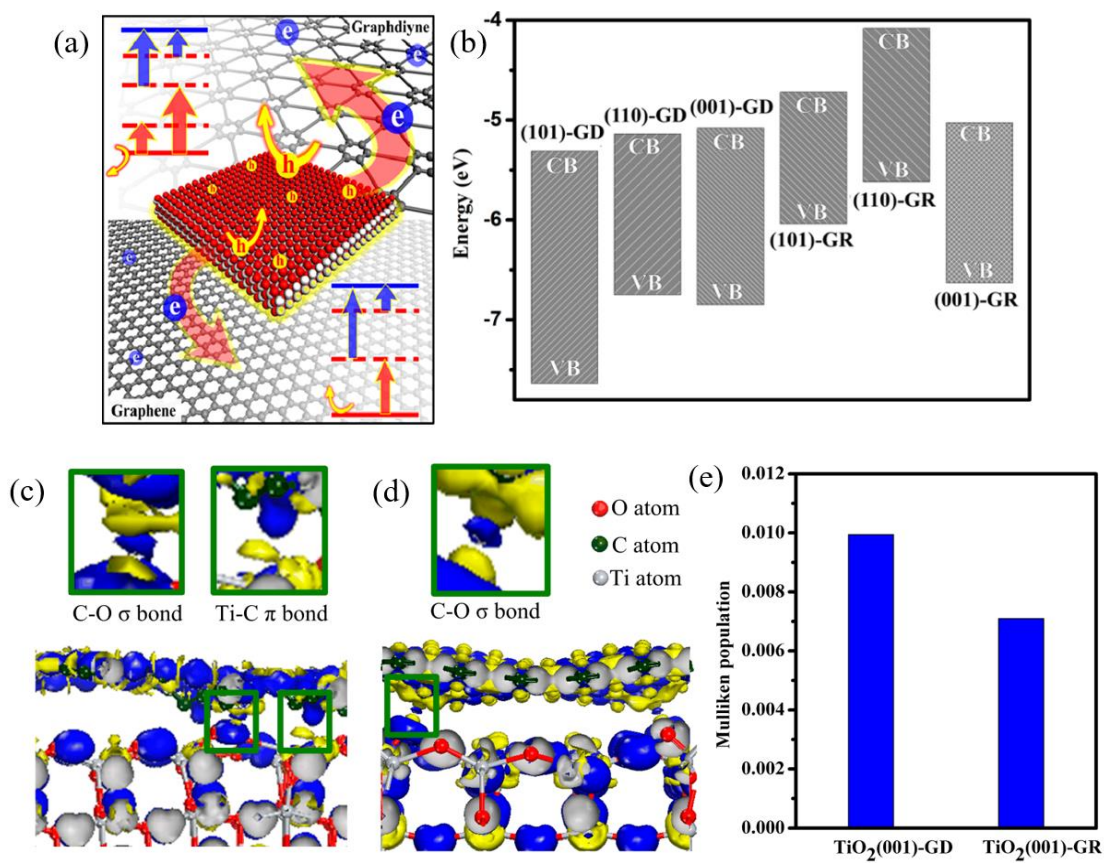

Figure 22(a). Photodegradation diagram of $TiO_2$ composite. (b). The conduction and valence band diagrams of different $TiO_2$ composites. Electron diagram of composite materials contact surface: (c). $TiO_2$-GDY, (d) $TiO_2$-GR. (e). Mulliken charge of different composite materials. Reproduced with permission from ref. [64]. Copyright 2013 American Chemical Society.

ZnO is a common chemical additive, which energy band gap and exciton binding energy are large. ZnO has a relatively high electron mobility, so it has been widely used in the semiconductor field in recent years. Under the excitation of light, the photocatalytic efficiency of ZnO is higher than the photocatalytic efficiency of $TiO_2$ in the same situation [171]. The photocatalytic performance of the composite material formed by ZnO and GDY is greater than that of pure ZnO. The reaction efficiency of pure ZnO material is 42%, while the reaction efficiency of GD-ZnO material is 26% higher than the former. GDY has good electrical conductivity, and this property can be combined with the highly active carriers of ZnO. GDY can obtain the electrons needed for the reaction from ZnO. These electrons react with oxygen molecules to achieve the function of decomposing pollutants.

GDY is oxidized to form GDY oxide, which increased oxygen-containing functional groups make the properties more active and can be used as a surfactant. Researchers mix silver, silver bromide and GDY to remove the pollution of azo dyes such as methyl orange [172]. When GDY oxide is added to this mixture, the photocatalytic degradation efficiency of the dye will increase. In summary, GDY is widely used in photocatalysis field.

$Ag_3PO_4$ is an emerging photocatalyst, when the wavelength of light is greater than 420nm, $Ag_3PO_4$ shows higher photocatalytic performance. However, due to the photocorrosion phenomenon in the photocatalytic process, this reaction will reduce the activity of photocatalysis. In order to solve this problem, the researchers combined $Ag_3PO_4$ and GDY to create a new Z-scheme, which has 12.2 times the oxygen production capacity of pure $Ag_3PO_4$ [173]. To further study the principle of the mixture of $Ag_3PO_4$ and GDY as a photocatalyst, the researchers added $Ag_3PO_4$-GDY catalyst to methylene blue. It has been found through experiments that GDY, as a hydrophobic nanostructure, can promote the transport of electrons in silver phosphate and convert positrons to oxides. This is the main reason for the high efficiency of $Ag_3PO_4$-GDY as a photocatalyst [174]. In addition to being a photocatalyst for removing pollution, GDY can also be used in photocatalytic hydrogen production systems. The Cds/GD structure containing 2.2% GDY can increase the photocatalytic efficiency by 2.6 times [175].

## 4.2 Electrocatalyst

Electrocatalysis refers to a catalysis that accelerates the transfer of charge between an electrode and an electrolyte. It is mainly used in the treatment of organic sewage and the degradation of chromium-containing wastewater. Two-dimensional planar grid materials such as GDY and carbon nanofibers are widely used as carriers for electrocatalysts. The mixing of butadiene units and GDY materials can accelerate the catalytic and redox reaction [65,176-178]. Platinum-based electrocatalysts are used in traditional alkaline fuel cells to improve the reaction activity, but such catalysts are relatively expensive and have certain toxicity. In order to find a substitute with excellent performance, the researchers doped GDY and nitrogen together to replace the platinum catalyst, as shown in Figure 23[65]. Compared with traditional platinum catalysts, nitrogen-doped GDY as an electrocatalyst has better stability and tolerance. After the nitrogen and carbon atoms are in contact, the latter will acquire positrons which increase the electron transmission speed. Nitrogen-doped GDY has great catalytic performance in alkaline fuel cells, but due to the poor reserve reaction effect of nitrogen in the redox process, the other electrical properties of nitrogen-doped GDY are still lower than traditional Pt/C electrocatalysts [179]. In order to further improve the electrical properties of nitrogen-doped GDY, researchers tried to dope nitrogen and fluorine together in GDY. The experiment proves that the N-F doped GDY exhibits the same level electrical properties as the traditional Pt/C, including electrocatalytic activity, starting potential and limiting current density, as shown in Figure 24[180]. For the purpose of understanding the nature of GDY, the researchers conducted calculations through the first-principles. The results of the study indicate that changes in the concentration of nitrogen will affect the optical and electrical properties of GDY. The modified NGDY has a very broad spectral range, which proves that NGDY has great potential in the application of electrocatalysts [181].

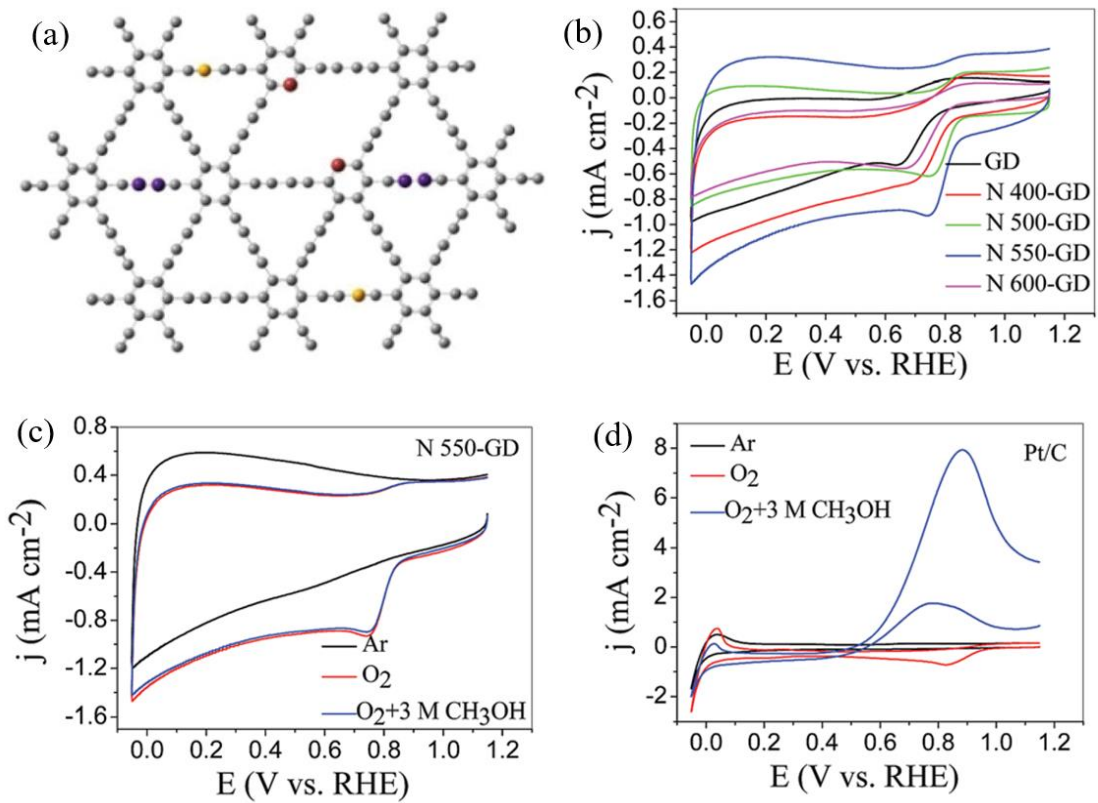

Figure 23. (a). Structure diagram of nitrogen-doped GDY. Among them, gray ball, yellow ball, purple ball and red ball respectively represent carbon atom, imine 1n, imine 2n and pyridinic nitrogen atom. (b). CV curves of GDY and nitrogen-doped GDY in 0.1M KOH solution saturated with oxygen. (c). CV curve of nitrogen-doped GDY in 0.1Mkoh solution saturated with argon. (d). CV curve of traditional platinum catalyst in 0.1M KOH solution saturated with argon. Reproduced with permission from ref. [65]. Copyright 2014 Royal Society of Chemistry.

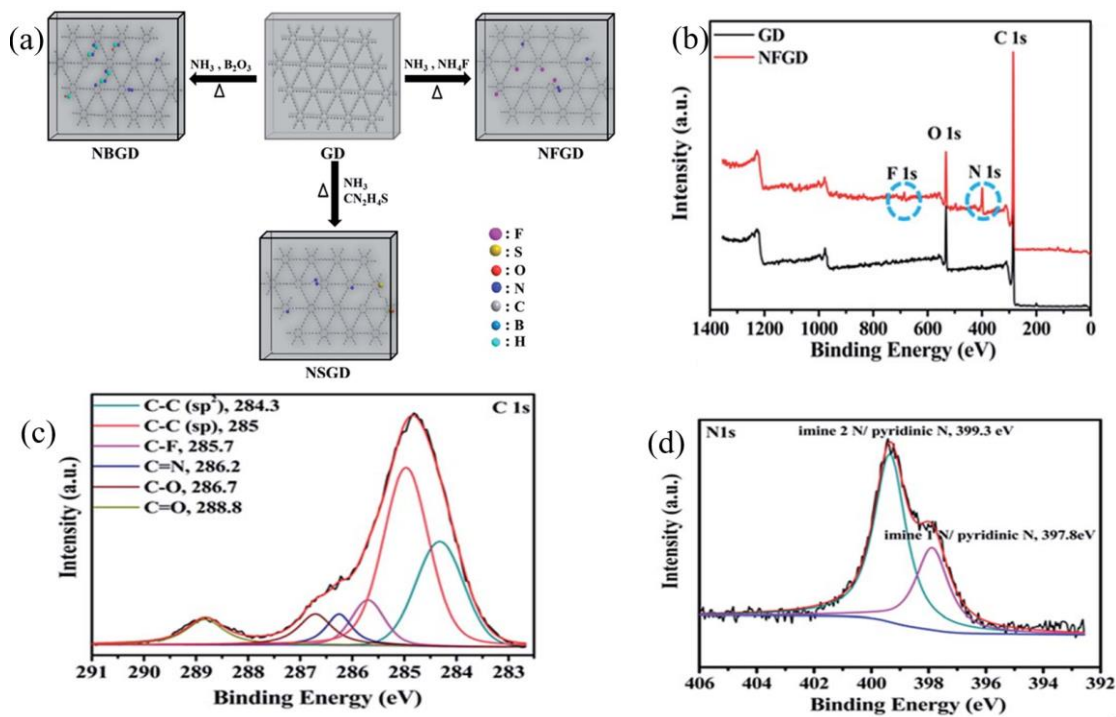

Figure 24. (a). schematic diagram of different elements doped GDY. (b). XPS spectrum measurement of GDY and NFGDY. (c). NFGD XPS C1s spectrum. (d). NFGD XPS N1s spectrum. Reproduced with permission from ref. [180]. Copyright 2016 Royal Society of Chemistry.

When N is doped into GDY, the hybrid carbon atoms in GDY will be replaced by N atoms. N atoms have a great influence on GDY materials, as shown in Figure 25 below [182]. This material has similar catalytic activity to traditional catalysts in alkaline solution and this material is very stable. How GNDC plays a role in the redox reaction is not clear. The researchers conducted a redox reaction experiment with the new GDY material for finding the site of the redox reaction. The results show that the carbon atom near pyridinic N is the main site for NGDY to participate in the catalysis [183]. The main way of doping N into GDY is annealing, at which time impurities in GDY are removed. In this way, the electrical performance and catalytic activity of GDY will be improved.

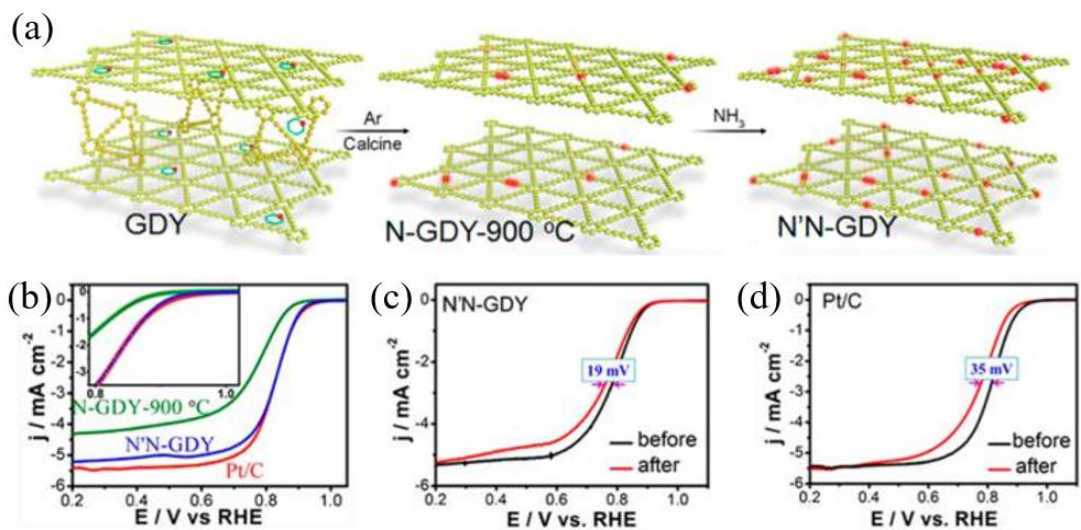

Figure 25. (a). NGDY synthesis process diagram. (b). Under the same environment, N-GDY-900° and Pc/T linear scanning voltammetry curve. (c). Under the same environment, N'N-GDY° and Pc/T linear sweep voltammetry curve. (d). Under the same environment, two Pc/T linear sweep voltammetry curves. Reproduced with permission from ref. [182]. Copyright 2017 American Chemical Society.

The electrocatalyst hydrogen evolution reaction (HER) is an important way for the industry to produce hydrogen cheaply. The use of GDY can effectively improve the efficiency of hydrogen evolution reaction. The researchers used GDY as the outer layer of the catalyst and copper as the core (Cu@GD NA/CF), and obtained the nanowire array as the catalyst, as shown in Figure 26[184]. Electrocatalytic reaction with Cu foam materials in $H_2SO_4$ with a concentration of 0.5M can be found that the catalyst activity is higher than the previous standard and higher than many non-noble metal catalysts. This phenomenon is mainly due to the doping of copper in GDY, which greatly reduces the impurities of the dopant. This method has the advantages of low cost and convenient preparation, and it has great prospects in industrial hydrogen production.

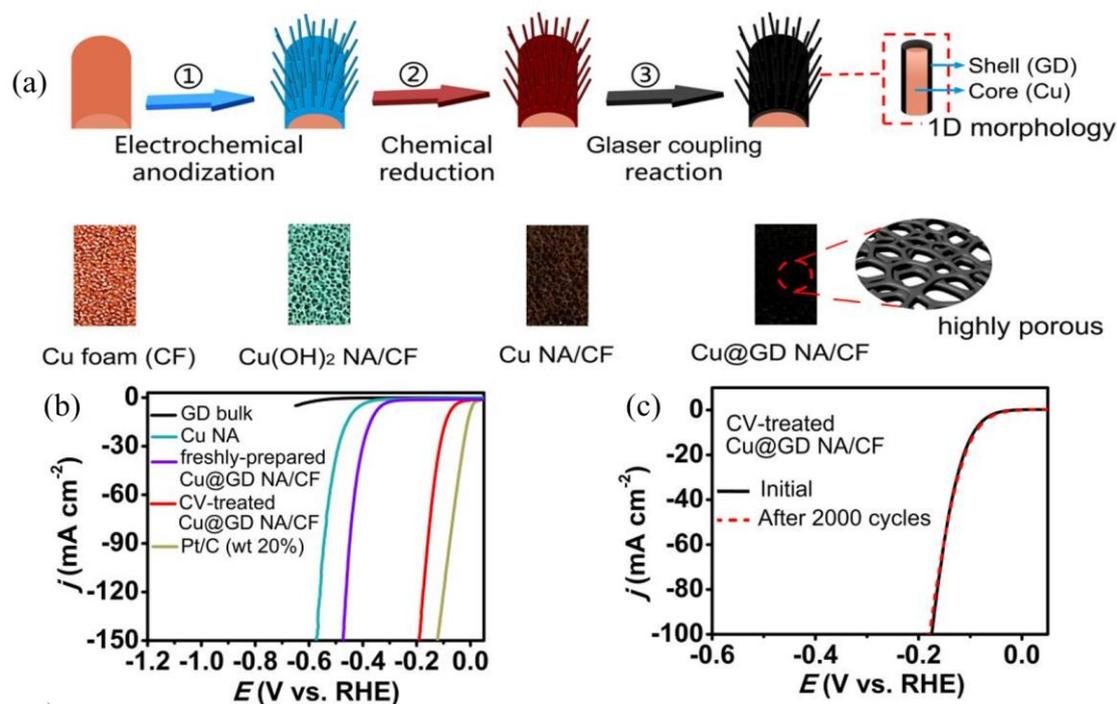

Figure 26. (a). Preparation process of Cu@GD NA/CF and pictures of test materials. (b). HER polarization curves of different materials. (c). HER polarization curves when Cu@GD NA/CF was subjected to 2000 times at 0-0.7V. Reproduced with permission from ref. [184]. Copyright 2016 Elsevier.

Cobalt can be doped into GDY and used as an electrocatalyst. The GDY nanomaterial is placed inside the cobalt element, and the outermost layer is made of carbon material doped N (CoNC/GD). When CoNC/GD is used as a catalyst, due to its excellent physical and chemical properties, it can accelerate the electron transfer of the electrode material and improve the catalyst activity, as shown in Figure 27[185]. It has been found through experiments that CoNC/GD also has high catalytic activity in HER. At the same time, because CoNC/GD is a non-precious metal material, the stability is relatively high. The durability of CoNC/GDis greater than that of commercial Pc/T. For example, after 38000 cycles of CoNC/GD in acidic conditions, the durability of the CoNC/GD material is higher than the durability of commercial Pt/C(10 wt %) under this condition.

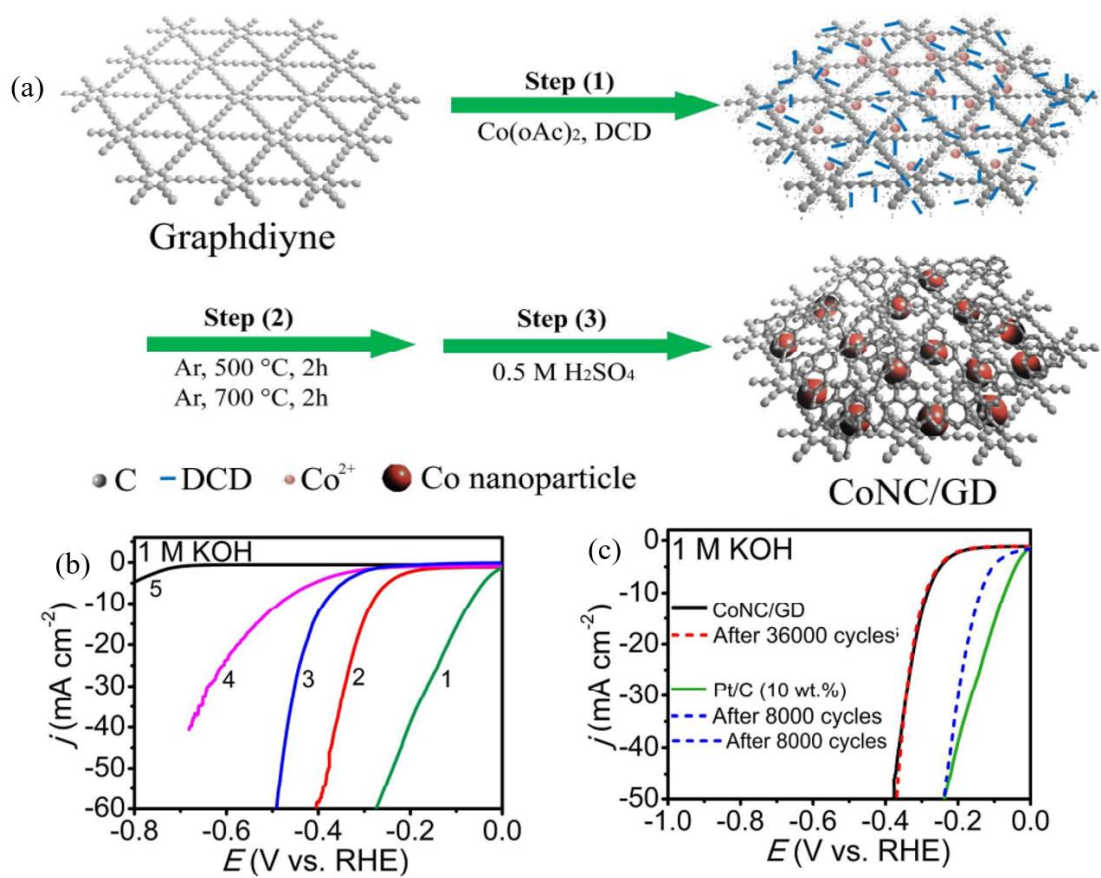

Figure 27. (a). CoNC/GD material preparation flow chart. (b). CoNC/GD HER polarization curve. (c). CoNC/GD HER polarization curve before and after a certain cycle. Reproduced with permission from ref. [185]. Copyright 2016 American Chemical Society.

Monoatomic transition metal catalysts have significantly different physicochemical properties relative to the bulk structure, and have attracted extensive attention from researchers in recent years. In order to fix the transition metal atoms on the carbon material, Xue et al. proposed a fast and controllable method to synthesize Ni/GDY and Fe/GDY[71]. As shown in Figure 28(a), GDY material is generated by cross-coupling reaction on carbon cloth material. Then the monoatomic metal anchored on GDY is formed by metal ions through redox reaction. Finally, Ni and Fe atoms can be fixed at different positions on GDY. As shown in Figure 28(b), the left picture shows the possible locations of metal atoms, and the right picture shows the best positions obtained through research. The catalytic activity of this signal-atom catalyst is significantly higher than that of traditional Pt/C catalyst. As shown in Figure 28(c), when the overpotential is 0.05V, the single-atom catalyst activity is more than ten times that of Pt/C catalyst. The overcharge is increased to 0.2V, and the catalytic activity of Fe/GDY is much greater than Pt/C. For the purpose of proving that the experimental results are qualified, the researchers used an electron microscope to take images of the manufactured materials, as shown in Figure 28(d)-(g). Figure 28(d) and Figure 28(e) show that the Ni atoms on the GDY are evenly arranged. Figure 28(f) and Figure 28(g) shows that the Fe atoms are uniform distribute. Subsequently, Xue et al. combined Pt atoms with GDY to produce two Pt single-atom catalyst. In the hydrogen evolution reaction , the catalytic activity is more than 20 times that of the traditional catalyst [186].

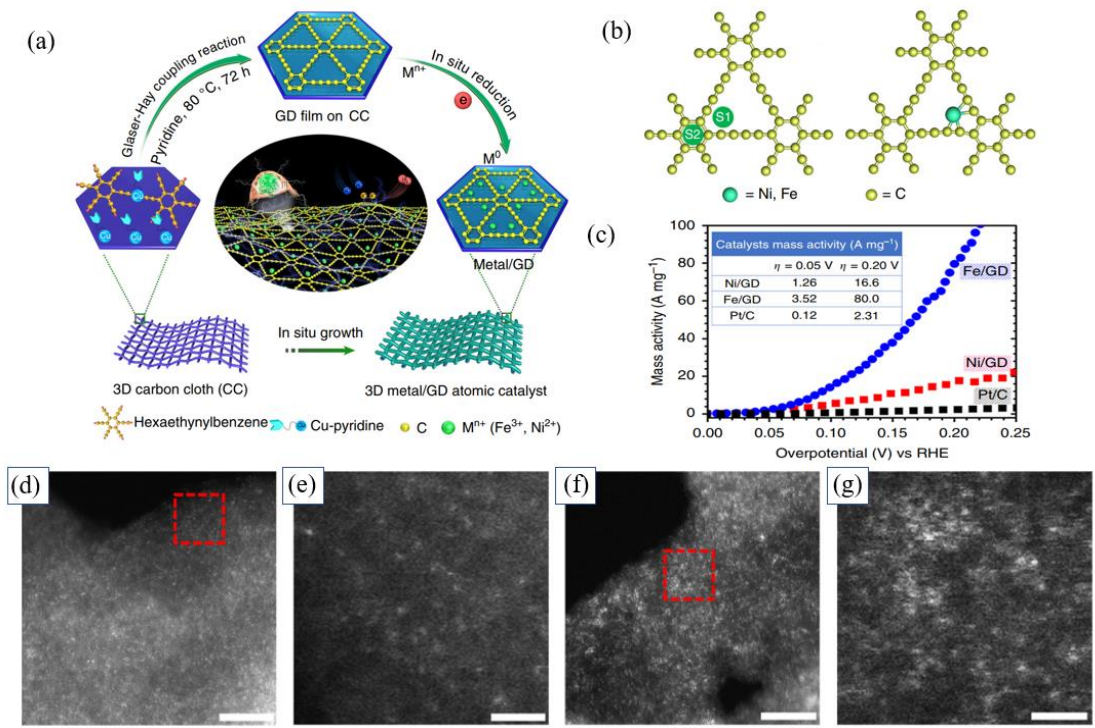

Figure 28(a). Flow chart for preparing single atom catalyst. (b). The position of metal atoms on GDY. (c). The curve of catalyst activity at different overpotentials. (d)(e). Enlarged image of Ni/GDY. (f)(g). Enlarged image of Fe/GDY. Reproduced with permission from ref. [187]. Copyright 2018 Springer Nature.

In recent years, lots of researchers have used GDY instead of metal as a support for electrocatalysts, due to its stable chemical properties. GDY is not easy to be corroded by the solution in the chemical reaction, and it can enhance the catalytic activity and durability[188]. $NiCo_2S_4$ nanowires are mounted on GDY foam substrate to form a dual-function catalyst (NW/GDF), which acts on the oxygen evolution and hydrogen evolution reactions, as shown in Figure 29. This NW/GDF can work at a lower voltage and perform water cracking in alkaline water. In addition, it can work for 140 hours in a 20mA/$cm^{-2}$ environment.

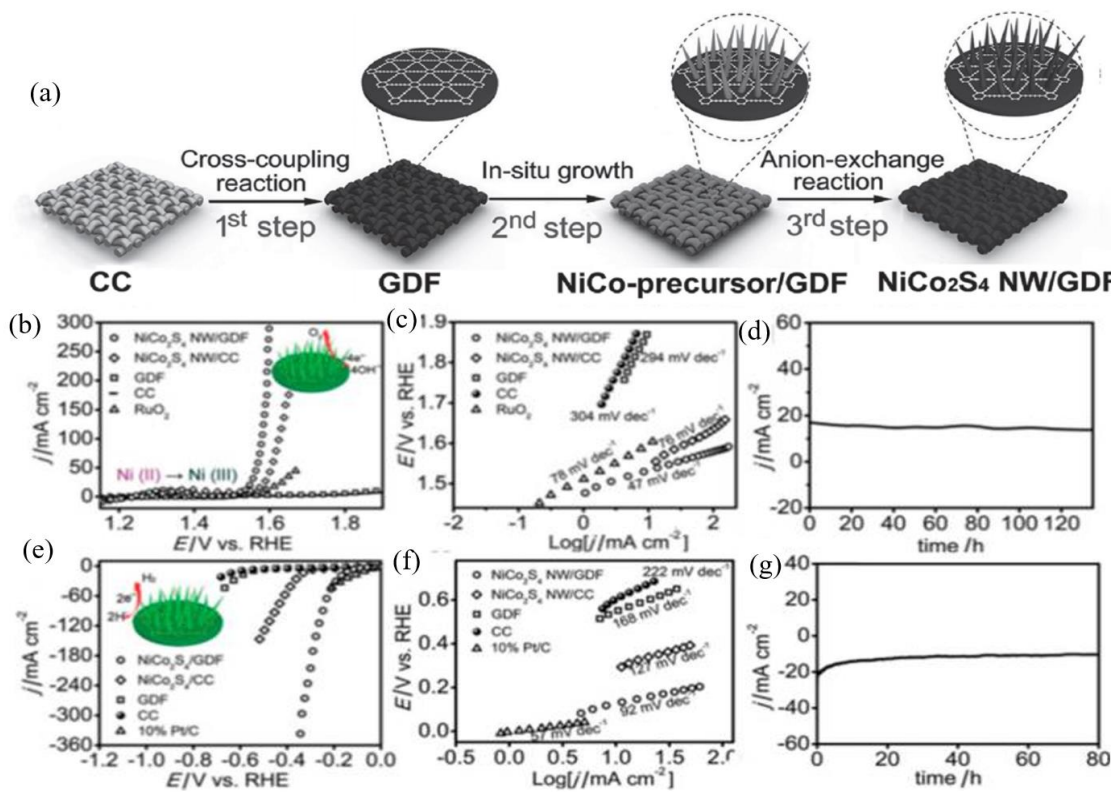

Figure 29(a). Flow chart of catalyst preparation using GDY foam and NiCo$_2$S$_4$ nanowires. (b),(e). Polarization curve of NW/GDF. (c),(f) Tafel slope of NW/GDF. (d) The stability test of oxygen evolution reaction. (g) The stability test of hydrogen evolution reaction. Reproduced with permission from ref. [188]. Copyright 2017 John Wiley and Sons.

A novel electrocatalyst for HER at any PH was synthesized by the combination of molybdenum disulfide and GDY (eGDY/MDS). S and Mo atoms are mounted on carbon cloth, which provides the catalyst with high activity and alkali resistance, as shown in Figure 30 [189]. A highly reactive reaction occurs between e-rich GDY and molybdenum disulfide, which accelerates the current transmission at the interface of the two substances, thereby accelerating the decomposition of water and the generation of hydrogen. The overpotential value of the catalyst in 0.5M sulfuric acid is 128mV, and the Tafel slope is 46mV dec$^{-1}$. In addition, the catalyst has better catalytic activity in alkaline solution than Pc/T, and the overpotential value in 1.0M KOH solution It is 99mV and the Tafel slope is 89mV dec$^{-1}$. This heterogeneous structure has a profound impact on the energy sector.

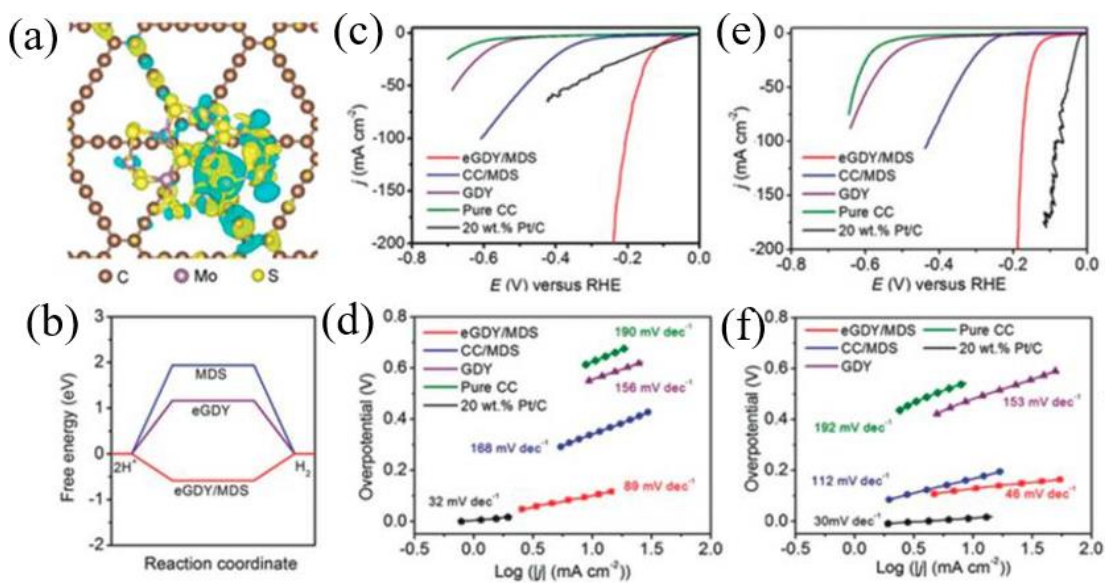

Figure 30 (a). eGDY/MDS charge map. (b). Material free energy diagram. (c)(d). Polarization curve and Tafel slope diagram of the material in 1.0MKOH solution. (e)(f). Polarization curve and Tafel slope diagram of the material in 0.5M sulfuric acid solution. Reproduced with permission from ref. [189]. Copyright 2018 John Wiley and Sons.

GDY and nickel-iron materials are combined into a mixed electrocatalyst (GDY/NiFe-LHD) under hydrothermal effect, and its overpotential is 260mv. (GDY/NiFe-LHD) catalyst has higher durability in HER in alkaline solution due to the excellent electrical properties of GDY [190,191]. $WS_2$ is often used as a petrochemical catalyst in industry which has high cracking performance, stable catalytic activity and long service life. In order to make the electrocatalyst acting on HER have better catalytic activity and stability in acidic solution, the researchers combined GDY and $WS_2$ to prepare a layered nanocatalyst(GD-$WS_2$ 2D-NH). Studies have shown that GD-$WS_2$ 2D-NH has an active built-in electric field that can enhance charge transfer in the catalyst. Its large-area defect-rich structure reduces the initial potential of HER to 140mv, and the Tafel slope is 54mv per decade[192].

Most traditional ORE catalysts are expensive to produce and have low output. Co nanoparticles and GDY were synthesized into a new 3DCu@GDY/Co, in which GDY was used as a carrier, and Co nanoparticles were mounted on GDY [193]. The unique conjugated structure of GDY and metal ions are connected together for self-reaction, which makes the electrode have higher catalytic activity. At this point, the overpotential is about 0.3V. Experimental results show that Cu@GDY/Co has better market prospects in ORE.

The study found that adding nitrogen-doped graphdiyne (NGDY) to HER has a positive effect on the reaction [194]. Spectral data indicate that the strong interaction between NGDY and $MoS_2$ will occur, and the electron transmission between the two will be enhanced, as shown in Figure 31. The heterostructure of this electrocatalyst results in an overvoltage of 186mv and a Tafel slope of 63mv dec$^{-1}$, and the catalyst stability will be greatly enhanced, which is greater than that of all existing $MoS_2$ based electrocatalysts. The experimental results show that NGDY is a HER catalyst with excellent properties, which provides a direction for future research.

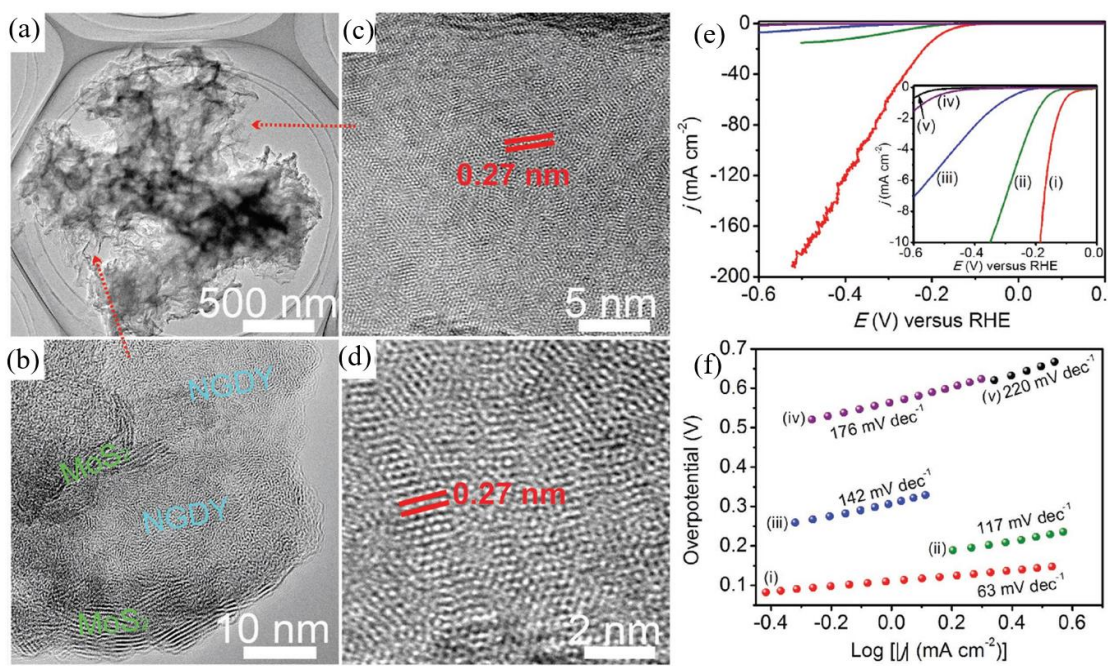

Figure 31(a). TEM image of the whole material. (b). HRTEM image of the whole material. (c). HRTEM image of MoS$_2$ part. (d). HRTEM image of GDY part. (e). Linear scanning voltammogram of several materials. (f). Tafel curve of several materials. Reproduced with permission from ref. [194]. Copyright 2018 John Wiley and Sons.

    Compared with traditional catalysts, single-atom catalysts have better selectivity and stability. The researchers combined molybdenum atoms and GDY to prepare Mo0/GDY atomic catalysts [195]. The catalyst is easy to prepare and the preparation process is flexible. Mo0/GDY is a dual function electrocatalyst, and it has great catalytic activity in the electrochemical nitrogen reduction reaction and HER. This catalyst laid the foundation for the further development of single-atom catalysts. In order to further deepen the research on the nitrogen reduction reaction by GDY, the researchers combined Co$_2$N and GDY to obtain an electrocatalyst (GDY/Co$_2$N) with high catalytic activity [196]. The chemical bond formed by cobalt and nitrogen is optimized by GDY. The experimental results show that in the acid solution, the amount of ammonia obtained can reach 219.72 μg h$^{-1}$ mg$_{cat.}$$^{-1}$. Figure 32 shows the working performance of GDY/Co$_2$N catalyst under different conditions. Under different potentials, this catalyst can show higher catalytic activity. This makes GDY/Co$_2$N become a very efficient electrocatalyst.

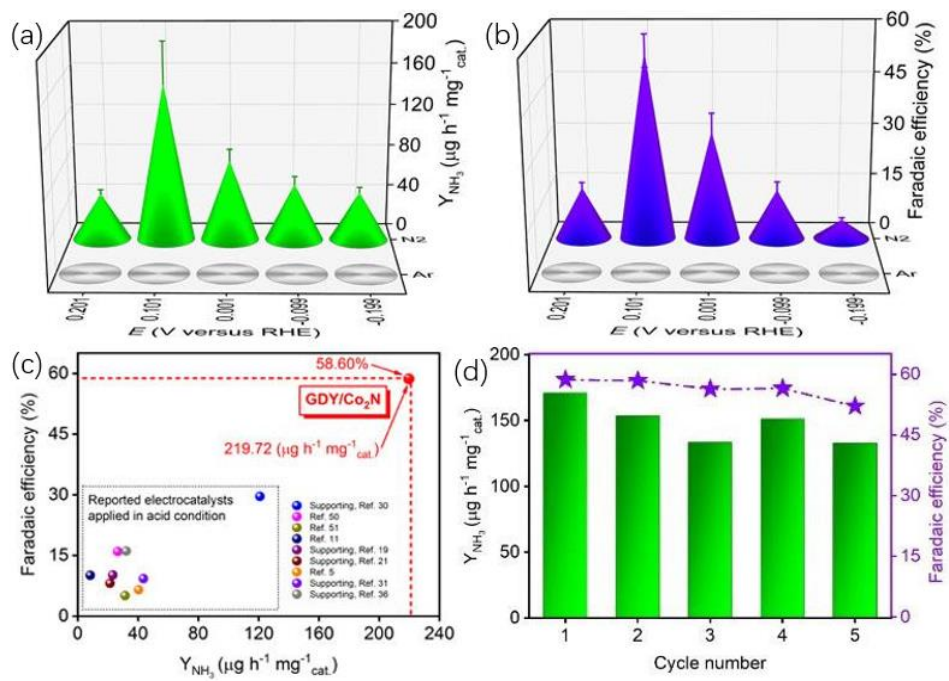

Figure 32. Performance of GDY/Co$_2$N catalyst for nitrogen reduction reaction in acidic solution. (a). Diagram of the output of NH$_3$ under different potentials. (b). Graph of faradaic efficiency of GDY/Co$_2$N at different potentials. (c). Comparison of GDY/Co$_2$N and other catalysts. (d). The cycle test chart of GDY/Co$_2$N when the potential is 0.101V. Reproduced with permission from ref.[196]. Copyright 2020 Wiley-Blackwell.

In order to study metal-free catalysts, a GDY-base metal-free nitrogen reduction catalyst was proposed, which is called crystalline fluorographdiyne (cFGDY)[197]. In neutral solution at room temperature, this catalyst can obtain NH$_3$ directly from nitrogen and water. As shown in Figure 33, cFGDY can maintain high catalytic performance under different conditions. The preparation of cFGDY is simple and the catalytic activity is perfect. It is an excellent metal-free catalyst that used to prepare ammonia.

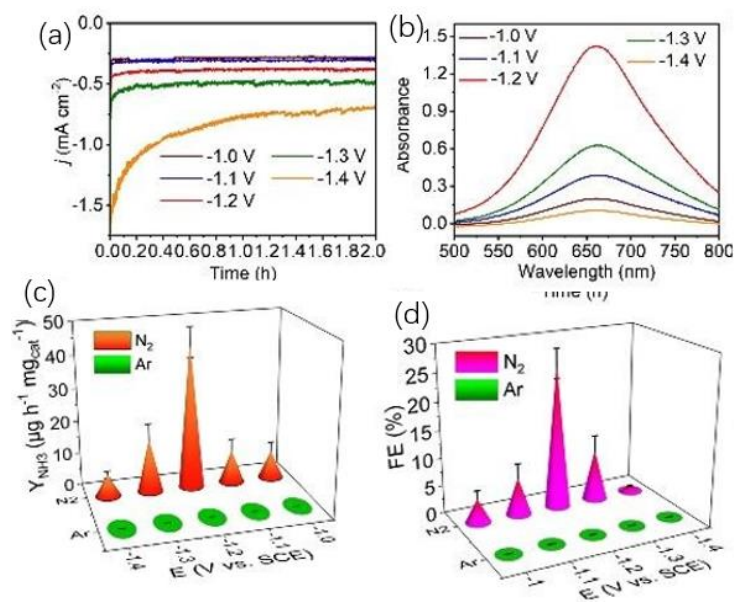

Figure 33. Catalytic performance of cFGDY under sufficient nitrogen. (a). Trend graph of current

density of cFGDY when potentials change. (b). The spectrum of the dielectric after the nitrogen reduction reaction. (c). Under the condition of sufficient nitrogen, output of $NH_3$ changes with different potentials. (d). Under the condition of sufficient nitrogen, faradaic efficiency changes with different potentials. Reproduced with permission from ref.[197]. Copyright 2020 The Royal Society of Chemistry.

### 4.3 Photoelectrochemical Water Splitting

$BiVO_4$ is one of the most widely studied photochemical catalysts. In order to improve the properties of $BiVO_4$ in the catalytic reaction, GDY nanowalls were prepared on the $BiVO_4$ substrate by copper coating catalysis, as shown in Figure 34 [198]. GDY/ $BiVO_4$ is mainly used for photochemical water splitting reaction. Due to the excellent electrical properties of GDY, the current density of GDY/ $BiVO_4$ is about doubled. Through this simple method, GDY can improve the photoelectric properties of many materials.

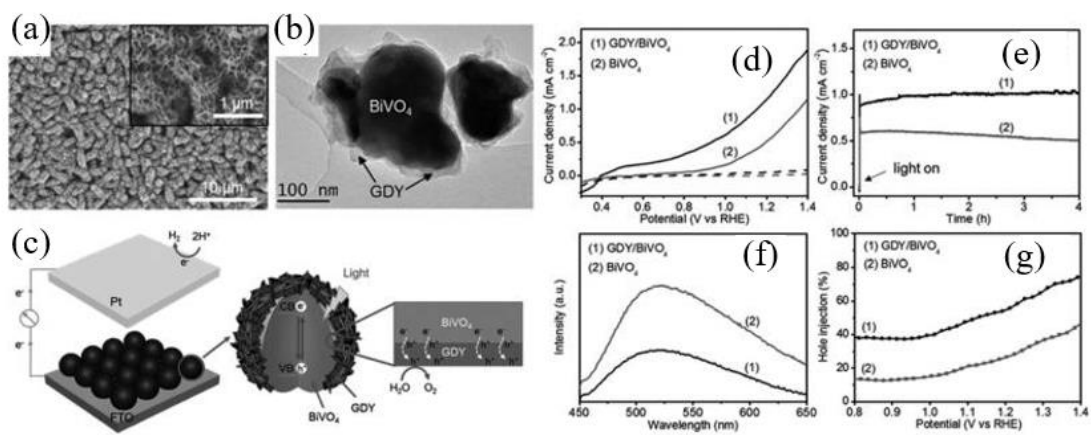

Figure 34. (a)(b). Microscope image of GDY/$BiVO_4$ nanostructure. (c). Experimental device for preparing GDY/$BiVO_4$. (d)(e). Current and voltage curves and J-t curve when $BiVO_4$ and GDY/ $BiVO_4$ are used as catalytic electrodes, respectively. (f)(g). Spectrum and hole injection yield of $BiVO_4$ and GDY/ $BiVO_4$. Reproduced with permission from ref.[198]. Copyright 2017 John Wiley and Sons.

With the continuous research on non-metallic elements, more and more non-metallic catalysts are used to replace traditional precious metal catalysts. $g-C_3N_4$ is a polymer semiconductor in which CN atoms hybridize to form a π-conjugated system. The semiconductor composed of g- $C_3N_4$ meets the thermodynamic requirements of photolysis water. The researchers made a catalyst composed of GDY and $g-C_3N_4$, which improved the problem of low hole mobility of the g- $C_3N_4$ semiconductor in the photolysis water reaction, as shown in Figure 35[199]. $g-C_3N_4$/GDY has an excellent 2D heterojunction structure, which increases its electron density and service life by three and seven times, respectively. If Pt atoms are added to $g-C_3N_4$/GDY, the photocatalytic performance will be further enhanced.

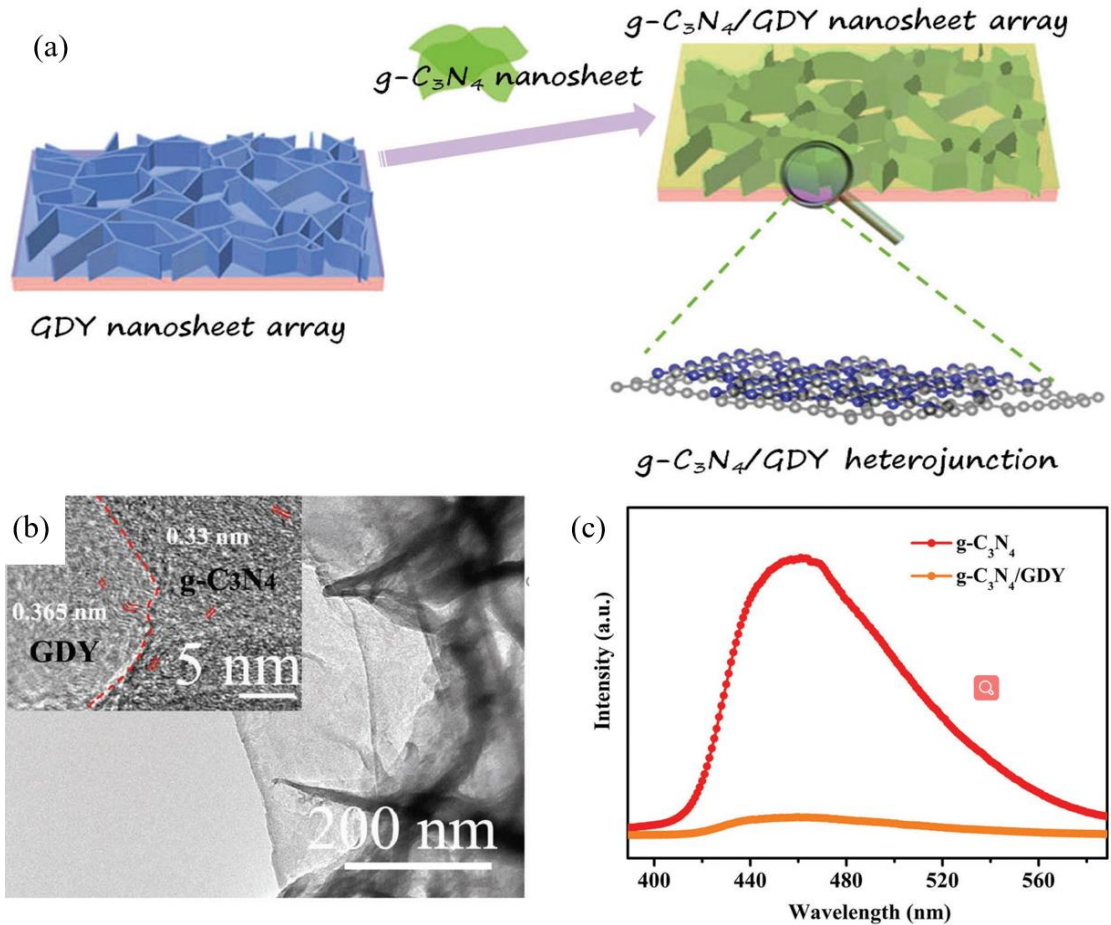

Figure 35(a). GDY layer and g-$C_3N_4$ combined structure diagram. (b). HRTEM image of g-$C_3N_4$/GDY. (c). Photoluminescence spectrum of g-$C_3N_4$/GDY. Reproduced with permission from ref. [199]. Copyright 2018 John Wiley and Sons.

- Nickel foam and GDY nanomaterials are made into an electrolyzed water catalyst (FeCH@GDY/NF), in which FeCH is wrapped by an ultra-thin GDY layer, as shown in Figure 36 (a) [200]. The rugged GDY film can greatly improve the durability of FeCH and increase the catalytic activity. Figure 36(b) and Figure 36(c) show that the lattice spacing of FeCH @ GDY / NF is smaller than that of FeCH, indicating that FeCH and GDY together intensify the reaction in the catalyst. Fe can increase the reaction surface area of the catalyst and enhance the current exchange between FeCH and GDY/NF. As shown in Figure 36(d) and Figure 36(f), the catalyst is very suitable for water-splitting reaction [201].

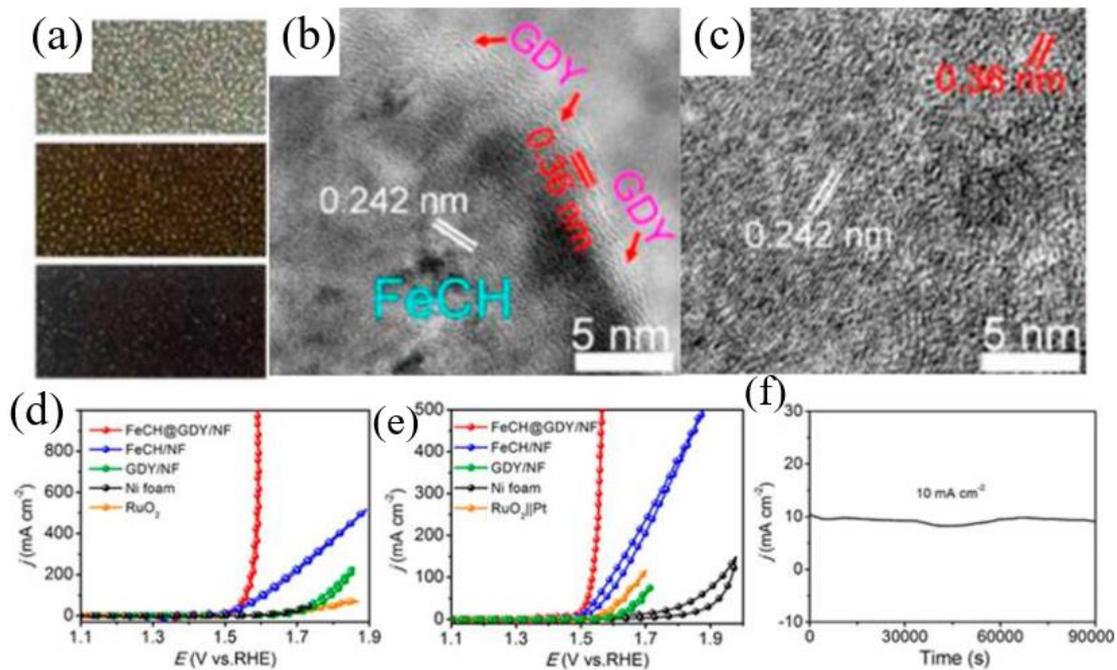

Figure 36(a). Images of NF, FeCH and FeCH@GDY/NF. (b)(c). HRTEM image of FeCH@GDY/NF. (d). CV curves of different catalyst materials in alkaline solution. (e). The CV curve of the material in the two-electrode experiment. (f). FeCH@GDY/NF current density and response time relationship curve. Reproduced with permission from ref.[200]. Copyright 2018 American Chemical Society.

The intercalation and exfoliation method induced by GDY was used to prepare a dual-functional water splitting catalyst [202]. This catalyst (e-ICLDH@GDY/NF) is prepared by double-layer hydrexide, iron-cobalt and GDY with intercalation andstriooing method. It shows excellent catalytic activity and stability in OER and HER, as shown in Figure 37. This is because GDY improves the catalytic site and increases the catalytic activity. In OER, when the overpotentials are 216, 249 and 278mV, the current densities are 10, 100 and 100mAcm$^{-2}$. In addition, when the overpotentials are 43, 215 and 256mV, the current densities are 10, 100 and 100mAcm$^{-2}$, respectively. The e-ICLDH@GDY/NF can be applied to water splitting reaction effectively. In summary, the GDY-based water splitting catalyst can effectively improve the performance of traditional catalysts due to its unique optical and electrical properties. The combination of different froms of GDY materials and matel, transition metals, non-metal materials etc. can produce wonderful changes. GDY has great potential in water splitting, hydrogen batteries and other applications [203,204].

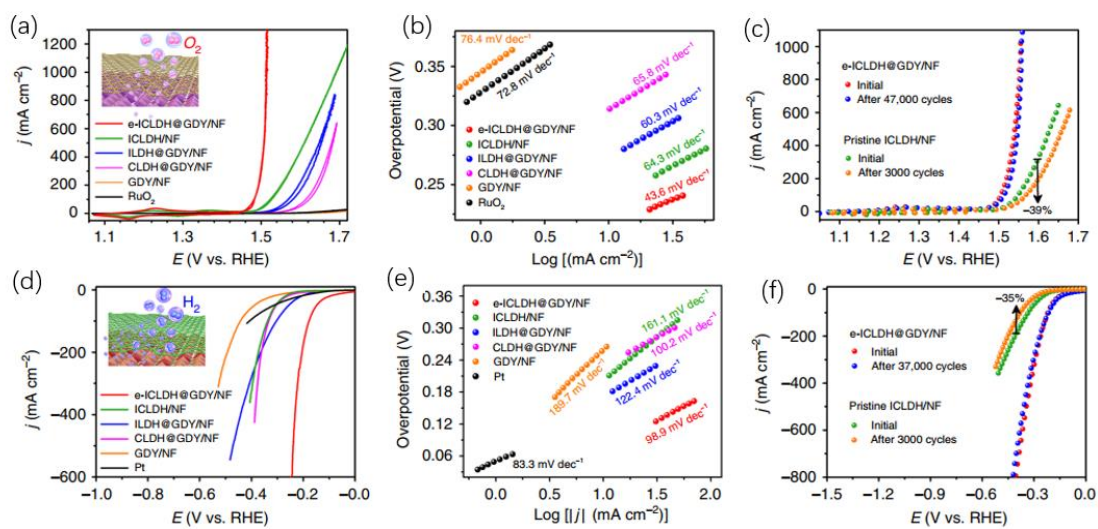

Figure 37. (a). CV curve and (b). Tafel slope of e-ICLDH@GDY/NF and several different catalysts in OER. (c). Polarization curves of e-ICLDH@GDY/NF catalyst after different cycles in OER. (d). CV curves and (e). Tafel slope of e-ICLDH@GDY/NF and several different catalysts in HER. (f). Polarization curves of e-ICLDH@GDY/NF catalyst after different cycles in HER. Reproduced with permission from ref.[202]. Copyright 2018 Nature Publishing Group.

### 4.4 Other Catalysis.

GDY can be oxidized to obtain GDYO which can be used as an excellent substrate for metal catalysts [205,206]. For example, through the redox reaction of GDY and $PdCl_4^{2-}$, Pd is deposited on GDYO to form a Pd/GDYO nanocomposite material, which has a strong reducing ability to 4-nitrophenol, as shown in Figure 38[206]. Due to the relatively uniform distribution of Pd on GDYO, the catalytic activity of Pd/GDYO catalyst for 4-nitrophenol is higher than that of traditional catalysts such as Pd/C.

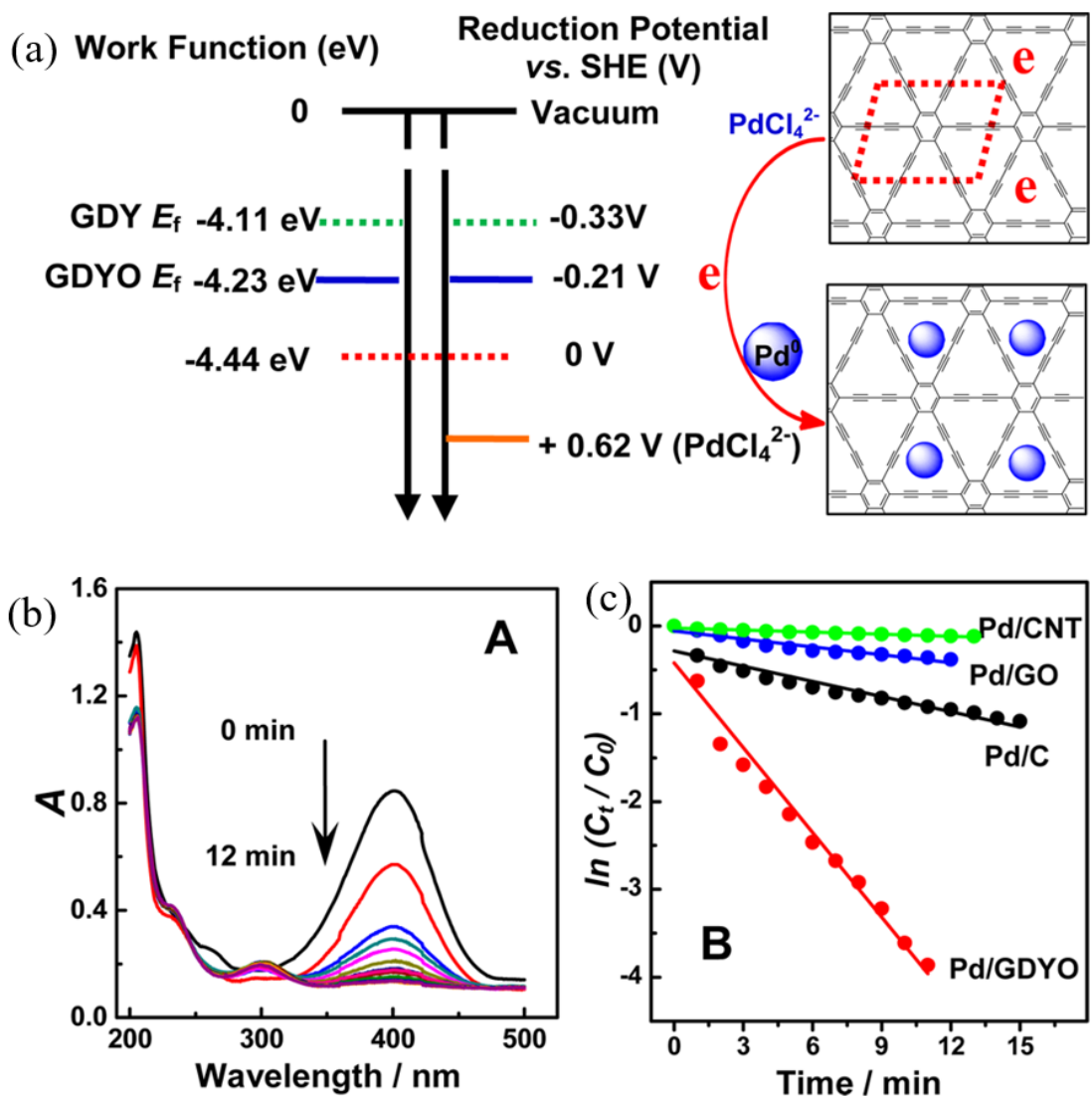

Figure 38. (a). Pd/GDYO preparation flow chart. (b). Ultraviolet absorption spectrum in the catalytic process of Pd/GDYO. (c). Reaction time function diagram of various Pd composite materials. Reproduced with permission from ref. [206]. Copyright 2015 American Chemical Society.

The adsorption capacity of precious metal (NM), Pd, Pt, Rh and Ir to GDY are measured separately. The geometric and electronic structures of four different atoms are compared. The experimental results show that Rh and Ir and GDY combination have the potential as metal catalysts [207]. Another study showed that GDY has a strong chemical adsorption capacity for nickel atoms, a typical chemical adsorption capacity for silver and copper atoms, and a weak physical adsorption capacity for zinc atoms. And GDY has better adsorption capacity for silver, copper and nickel than graphene[208].

In order to solve the increasing pollution caused by CO to the environment, how to oxidize CO more efficiently has become a challenge. The cost of traditional metal catalysts is higher, and the reaction environment requirements are stricter. Therefore, a method using GDY as a catalyst for oxidizing CO is proposed [48]. In this process, GDY absorbs oxygen molecules in the air, and the O-O band of $O_2$ is broken. After that, oxygen atoms and CO undergo an oxidation reaction to form $CO_2$. The experimental results show that GDY is a catalyst with excellent performance, low cost and no pollution. GDY can break the O-O bond in oxygen and oxidize CO into carbon dioxide molecules.

When Pt nanostructures are used as catalysts in fuel cells, the problem of losing catalytic performance occurs. A new catalyst synthesized by mixing GDY in Pt can solve this problem. The pore structure of GDY can effectively adsorb Pt particles and solve the cheap problem of pure Pt nanostructures in fuel cells [209-211]. The experimental results show that GDY can prevent the aggregation of metal catalysts and optimize the catalytic performance of the catalysts. The conversion of GDY and $Ag_{38}$ into $Ag_{38}$ GDY can improve the oxidation performance of the catalyst and reduce energy barrier to 0.26V [210].

The dehydrogenation reaction means that the hydride undergoes an oxidation reaction under the action of a catalyst, so that hydrogen atoms are detached from the compound. For this reason, the interaction between GDY and metal hydride has been studied as a catalyst for dehydrogenation reactions [212]. In addition, the researchers found that there is a stronger force between the light metal doped with hydride and GD or GP, which indicates that the stability of the chemical bong formed by the light metal and H atom will become easy to be broken, and this is beneficial to the reaction rate of dehydrogenation reaction. The experimental results show that GDY can significantly reduce the hydrogen removal energy, making the process of oxidizing hydride easier. In addition, the porous structure of GDY allows GDY-light metal mixtures to be used as hydrogen storage media.

The catalysts for the hydrogenation reaction mainly include transition metals, transition metal salts and transition metal complexes. Pt nanoparticles were used to synthesize a hydrogenation catalyst (Pt NPs) with GDY substrate, as shown in Figure 39 [213]. Figure 39(a) shows the catalytic mechanism of the Pt-GDY catalyst. In Figure 39(b), 1-5 respectively represent different reactants, and the researchers compared the catalytic efficiency of the proposed Pt-GDY and Pt-C catalyst. It can be seen that the Pt-GDY catalyst has a higher conversion efficiency than the traditional catalyst. Using GDY as a substrate can effectively increase the current density of Pt NPs, which makes the catalytic performance of Pt NPs higher than that of traditional catalysts such as Raney nickel or cobalt.

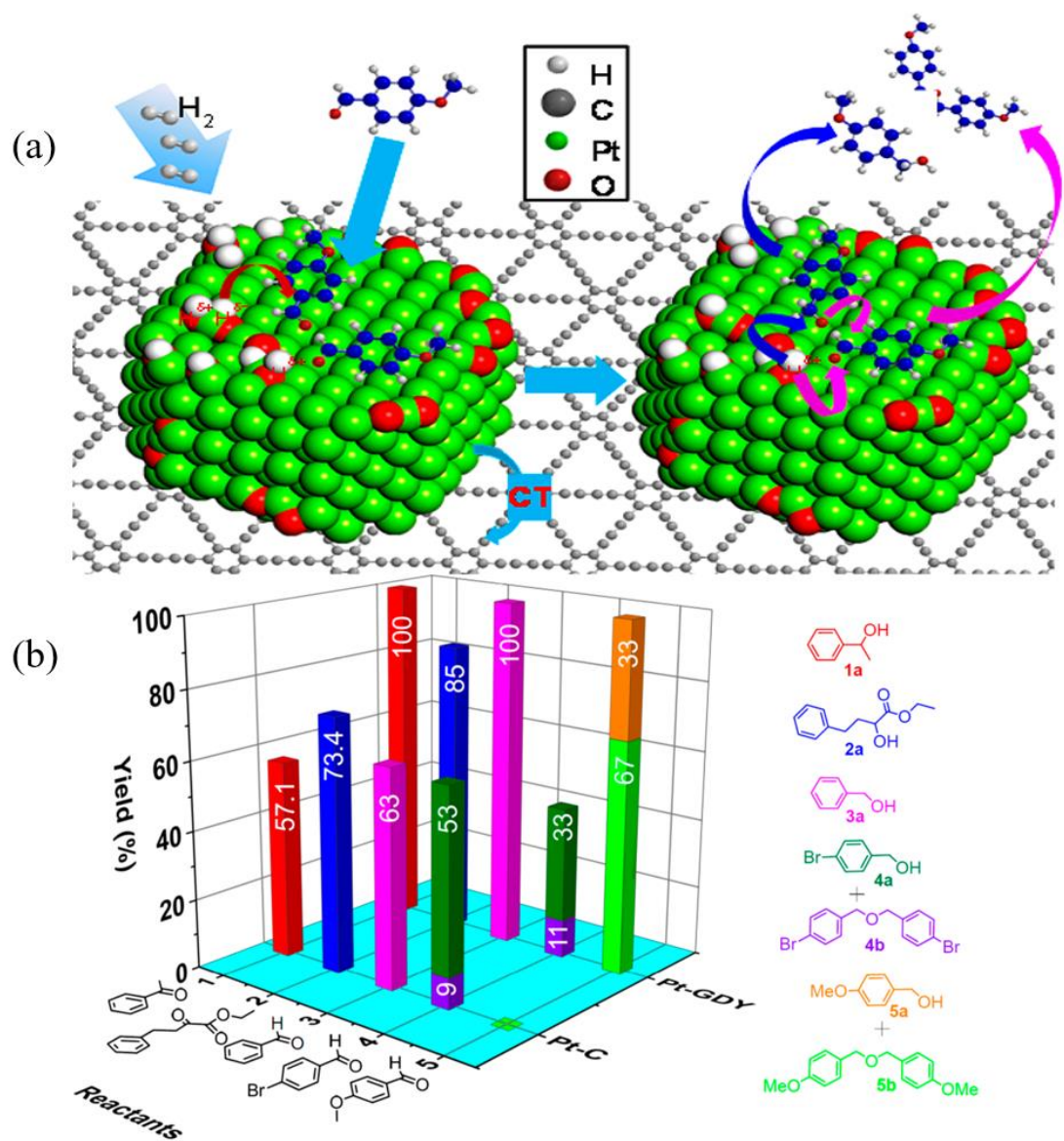

Figure 39. (a). Schematic diagram of hydrogenation catalysis. (b). Comparison of the efficiency of Pt NPs and Pt/C for 4h hydrogenation. Reproduced with permission from ref. [213]. Copyright 2018 American Chemical Socitey.

## 5. Conclusions and prospects

    In 1968, Baughman first discovered the concept of GDY materials through calculation, which is a 2D planar material with high π-conjugation. GDY is an allotrope of carbon formed by sp- and $sp^2$- hybridization linked by a diyne bond and a benzene ring. GDY materials contain rich carbon chemical bonds, a large conjugated system and uniform pore distribution, which makes GDY have excellent physical and chemical properties. The C-C structure of GDY can make the generation of nanomaterials more convenient. Its 2D material layer can form a 3D structure by van der Waals force, in which carbon atoms can be evenly distributed. The traditional preparation method of GDY uses HEB as the precursor, copper foil as the substrate and catalyst, and a multi-layer GDY film can be obtained through a series of reaction. In addition, a variety of GDY nanostructured materials have been successfully prepared recently. GDY has representative mechanical, electrical, optical and other characteristics, which can be applied to a variety of catalysts. Compared with

traditional catalysts, GDY can greatly improve catalytic efficiency and durability of catalyst materials. Such as, photocatalysis, electrocatalysis, water splitting catalysis and redox reactions.

At present, the biggest challenge in the application of GDY is the preparation problem. Even though there are many kinds of preparation methods for GDY proposed by researchers, lwo-cost and large-scale preparation techniques are still immature. This problem makes GDY play a relatively small role in practical applications. As summarized in this review, a method using HEB and copper foil can only obtain multi-layer GDY materials, not single-layer materials. Another method by controlling the GDY powder content and heating position can obtain different layers of GDY through the gas-liquid-solid process. Although this way can obtain signal-layer GDY, the interval between different layers is only 0.36mm, it is hard to separate them. In addition, some researchers used Ag as the substrate to prepare GDY, a single layer GDY can be obtained, but this method can only prepare a small amount GDY. In short, low-cost mass production of GDY is one of the topics of future research,

The detection method of GDY materials is also a key research issue. The chemical bond structure of GDY materials is different from the double bond structure of graphene. GDY also contain triple bond structure, which make the structure of GDY material not fixed. The researchers calculated three GDY structures through density functional theory. Some of GDY materials are hexagonal structures, while the rest are rectangular lattice structures. Different forms of GDY will show different properties in application. At present, researchers can use Raman spectroscopy to detect the presence of diacetylene bonds in GDY materials, but this method cannot specifically understand the proportion of other chemical bonds in GDY. This problem makes it different to discover the properties of different forms of GDY, resulting in the inability of GDY materials to be better applied in the practice. Therefore, the effective detection technology of GDY materials is also a subject that should be focused in the future.

The last issue that needs attention is the transfer method of GDY. The single-layer GDY film prepared by various methods is easily damaged. At present, the use of a conventional container (PMMA) can effectively prevent the integrity of the film from being damaged, but the impurities remaining in the container will corrode the GDY film during the transfer process. This problem will change the electrical properties of GDY film. Hydrofluoric acid is also added in the transfer process, which reduces the peak of the GDY Raman spectrum. In addition, hydrofluoric acid may have other unknown effects on GDY materials. Recently, the method of preparing GDY film directly at the experimental location without the transfer process has been proposed, but the process of generating GDY may have an impact on other reactive materials. Therefore, a clean GDY transfer method is essential for the development of GDY. We hope that this review can serve as a guide to the development of GDY.

## Conflicts of interest

There are no conflicts to declare.

## Acknowledgements

This work was partially supported by the National Natural Science Foundation of China (CN) (Grant Nos. 61605016; 6180021914).